\documentclass[runningheads,fleqn]{svmult}

\usepackage{makeidx}   

\usepackage{graphicx}  
                       
\usepackage[nocompress]{cite}                       
                       
\usepackage{subeqnar}  
\usepackage{multicol}  
\usepackage{physmult}  
\makeindex             

\newcommand{\greeksym}[1]{{\usefont{U}{psy}{m}{n}#1}}
\newcommand{\umu}{\mbox{\greeksym{m}}}

\newcommand{\nm}{\,\mathrm{nm}}
\newcommand{\mm}{\,\mathrm{mm}}

\newcommand{\mW}{\,\mathrm{mW}}

\begin{document}
\title*{Hybrid Quantum Nanophotonics – Interfacing Color Center in Nanodiamonds with $\mathrm{Si}_3\mathrm{N}_4$-Photonics}
\toctitle{Hybrid Quantum Nanophotonics}
\titlerunning{Hybrid Quantum Photonics}
%
\author{Alexander Kubanek*
\and Anna P. Ovvyan*
\and Lukas Antoniuk
\and Niklas Lettner
\and Wolfram H. P. Pernice \\
*corresponding authors: \\
alexander.kubanek@uni-ulm.de\\
ovvyan@uni-muenster.de}

\authorrunning{Alexander Kubanek et al.}
%
%

\maketitle              

\section{Abstract}

This chapter covers recent developments in the field of hybrid quantum photonics based on color centers in nanodiamonds and $\mathrm{Si}_3\mathrm{N}_4$-photonics towards a technology platform with applications in quantum information processing and quantum information distribution. The methodological approach can be divided in three main tasks. First, the fabrication and optimization of $\mathrm{Si}_3\mathrm{N}_4$-photonics. Second, the creation, characterization and control of color centers in nanodiamonds. Third, the assembly of hybrid quantum photonics by integrating the nanodiamonds into the photonic structures. One focus will be the efficient interfacing of the color centers done by optimizing the optical coupling. The chapter describes recent progress in all three steps and  summarizes the established hybrid platform. We believe, that the hybrid approach provides a promising path to realize quantum photonic applications, such as quantum networks or quantum repeaters, in the near future.

\section{Introduction}

Optics and photonics nowadays make a major contribution to leading edge technologies. As an example, fiber-optic communication system transfer information over long distances such as transatlantic connection and provide high bandwidth, low optical loss (from $0.22\,\mathrm{dB}/\mathrm{km}$), almost no crosstalk and protection against outside electromagnetic waves. Therefore, effective speed of data transmission with rates up to 11 Tbits/s or faster became possible \cite{winzerFiberopticTransmissionNetworking2018,polettiHighcapacityFibreopticCommunications2013}. Analogous, integrated photonic circuits (IPC) \cite{bogdanovMaterialPlatformsIntegrated2017} and nanophotonics \cite{kinseyExaminingNanophotonicsIntegrated2015} demonstrate strong potential to provide high-capacity information transmission, processing and communication on-chip. Operating the technology in the quantum regime \cite{murrayQuantumPhotonicsHybrid2015} could play the key role of an interruptive technology enabling a plethora of new applications and with the potential to herald a new era of communication \cite{paraisoPhotonicIntegratedQuantum2021}, computing and data security. \\
A field known as integrated quantum photonics enables to establish a technology based on sub-wavelength light-matter interaction where classic and quantum effects are conjoined through the utilization of nanophotonic building blocks. Main components of the integrated quantum photonic circuit include:\\
\begin{enumerate}
\item
Single photon emitter, ideally producing single photons on-demand; 
\item
Low-loss waveguiding, passive elements such as filters and beamsplitters; 
\item 
Reconfigurable active elements such as modulators and switches;
\item 
Cavities to enhance light-matter interaction such as photonic crystal (PhC) cavities or ring resonators; 
\item
Fast and efficient single-photon detectors. 
\end{enumerate} 
Integrated photonics enables the development of next-generation devices based on a technology that can withstand the demands of industrial production. However, the fabrication and control of quantum systems is at an early stage of the Technology Readiness Level (TRL).
The idea of combining IPCs, which are rapidly improving due to quickly evolving industry standards, with quantum optical systems of reduced complexity leverages the advantages of both worlds. Complex architectures can be realized by merging the strengths of the different systems. Operating the platform in the quantum regime opens the field of hybrid quantum photonics \cite{bensonAssemblyHybridPhotonic2011,kimHybridIntegrationMethods2020}. Hybrid quantum photonics holds the unique potential for a scalable technology with quantum properties of high-quality. Major challenges include the identification of the most promising systems to be combined and the establishment of a hybridization method that is efficient and salable for high-throughput production. The hybridization procedure is called quantum post-processing (QPP) in the following. This chapter focuses on a relatively new, yet rapidly evolving hybridization method utilizing color center in nanodiamonds (ND) to post-process silicon nitride ($\mathrm{Si}_3\mathrm{N}_4$)-photonics.

\section{Potential Pathways towards Hybrid Quantum Photonics}

A number of different strategies to build hybrid quantum devices based on color center in diamond were explored in the past decade. The methodologies can be divided in two main concepts. One, the \lq{}\lq{}structure to emitter" approach. Two, the \lq{}\lq{}emitter to structure" approach.

\subsection{\lq{}\lq{}Structure to Emitter" Approach}

The structure to emitter approach was realized in pioneering experiments by Barclay et al. \cite{barclayHybridPhotonicCrystal2009} and Englund et al.\cite{englundDeterministicCouplingSingle2010}. For example, the coupling between the mode of a PhC nanocavity and individual negatively-charged nitrogen-vacancy (NV) centers located close to the surface of bulk diamond, was realized by in situ lateral scanning of the nanocavity with respect to the location of the NV center \cite{englundDeterministicCouplingSingle2010}. An alternative realization of the structure to emitter approach builds on the fabricating of the photonic device around a preselected color center. For example, lithographic alignment is used to build an integrated platform by constructing plasmonic waveguides around precharacterized Germanium Vacancy center (GeV) in NDs in a deterministic fashion \cite{siampourUltrabrightSinglephotonEmission2020a}. The platform yields a 10-fold enhanced optical decay rate and unidirectional emission. Another very efficient assembly method includes all-silver nanopatch antennas, realized by silver nanocubes deposited on a silver substrate, where NDs are localized between in the area of strong field enhancement. For NV centers in NDs a 70-fold  enhancement of the fluorescence lifetime and a 90-fold intensity increase was demonstrated with a count rate of up to 35 million per second \cite{bogdanovUltrabrightRoomTemperatureSubNanosecond2018}. Integration of GeV centers in NDs into plasmonic modes of dielectric ridges atop colloidal silver crystals yields a 6-fold Purcell enhancement, $56\,\%$ coupling efficiency and $33\,\umu\mathrm{m}$ transmission length \cite{siampourOnchipExcitationSingle2018}. A $20\,\umu\mathrm{m}$ single photon propagation length, originating from NV fluorescence, was realized by building dielectric loaded surface plasmon polariton waveguides around the NDs using electron-beam lithography yielding 5-fold enhancement in the total decay rate and 58 \% coupling efficiency \cite{siampourNanofabricationPlasmonicCircuits2017}. And a chip-integrated cavity-coupled emission into propagating surface plasmon polariton modes enabled a 42-fold decay rate enhancement \cite{siampourChipintegratedPlasmonicCavityenhanced2017}. 

\subsection{\lq{}\lq{}Emitter to Structure" Approach}

The emitter to structure approach is an assembly method complementary to the structure to emitter approach. In this chapter, we focus on emitter to structure approaches involving the following steps:
\begin{enumerate}
\item 
The quantum emitters are prepared inside small-sized solid state hosts.
\item 
Independently, the photonics devices are optimized with pre-defined interaction zones.
\item 
In the last step, the emitter is placed in the interaction zone via nanomanipulation tools, a step which we label as QPP.
\end{enumerate}
In the past years, a number of different hybrid quantum photonic devices were realized following this procedure. Also nanostructured devices that concentrate electromagnetic fields in a region much smaller than the wavelength of light were interfaced with field enhancement of several orders of magnitude \cite{NanophotonicsSurfacePlasmons2006,gramotnevPlasmonicsDiffractionLimit2010}. Other techniques do not rely on nanomanipulation but instead utilize a charged ND surface to attach larger quantities of ND to, for example, tapered optical fibers \cite{vorobyovCouplingSingleNV2016}. A few examples of hybrid plasmonics and hybrid photonics based on color centers in ND and based on the emitter to structure approach are depicted in the following. However, it is beyond the scope of this chapter to give a complete overview. NDs containing NV centers were positioned in the interaction zone of PhC structures by using the pick and place technique \cite{schellScanningProbebasedPickandplace2011}. NV-center in  NDs were also coupled to surface plasmons in a controlled way leading to an enhanced photon emission rate \cite{huckControlledCouplingSingle2011} and further coupled to quantum plasmonic circuits \cite{huckCouplingSingleEmitters2016}.
When coupled to nanopatch antennas the emission rate of single NV center can be enhanced into an ultrabright regime with a sub-nanosecond emission \cite{bogdanovUltrabrightRoomTemperatureSubNanosecond2018}. NV \cite{andersenHybridPlasmonicBullseye2018} and silicon-vacancy (SiV) \cite{waltrichHighpuritySinglePhotons2021} center in NDs were placed in the center of bullseye antennas achieving a high-purity single photon source and enabling access via moderate-NA optics.  
An essential improvement towards an efficient integration of color centers in NDs into photonics is the usage of nanopockets. Placing a ND in a nanopocket within a PhC L3 cavity yields a robust integration and an up to 10-fold increased  Purcell factor as compared to evanescent coupling could be reached \cite{alagappanDiamondNanopocketNew2018}. Efficient integration of ND into silicon nitride photonics is proposed by using trenched nanobeam cavities. An optimized ratio between quality factor and mode volume can be achieved by a redistribution of the cavity modes originating from the presence of the ND \cite{alagappanPurcellEnhancementLight2020}.\\
Paths towards large-scale integration of individual color centers in NDs were explored and realized by means of lithographically positioned NV centers into photonic integrated circuits \cite{schrinnerIntegrationDiamondBasedQuantum2020}. As a result the lithographic positioning yields NDs in $70\,\%$ of the targeted locations with Purcell-enhanced coupling. Optically detected magnetic resonances (ODMR) was demonstrated while keeping simultaneous control over several NV center. \\
One of the most advanced and optimized procedures to date involves nanophotonic fabrication, for example of 1D PhC cavities, directly in diamond.
All-diamond nanophotonics was pioneered a decade ago \cite{babinecDiamondNanowireSinglephoton2010} and went through rapid progress since then to establish efficient spin-photon interfaces \cite{burekHighQualityfactorOptical2014}. Nowadays, the technology builds on well-developed all-diamond devices to realize quantum applications such as memory-enhanced quantum communication \cite{bhaskarExperimentalDemonstrationMemoryenhanced2020,sipahigilIntegratedDiamondNanophotonics2016}. In order to realize quantum computation applications, entanglement needs to be distributed among many nodes. Therefore, in a second fabrication step all-diamond microchiplets can be integrated into large-scale non-diamond photonics to realize wafer-scale devices. In pioneering work, defect free arrays with 128 channels of negatively-charged GeV- and SiV-center in aluminum nitride photonic integrated circuits were realized \cite{wanLargescaleIntegrationArtificial2020}.\\
\\
In this review, we focus on the emitter to structure approach and, in particular, on a process chain where no diamond nanofabrication is required. Instead, all nanofabrication is performed on classical photonics materials only, here in particular on $\mathrm{Si}_3\mathrm{N}_4$. The only requirement for the diamond host is on the size which needs to be small enough to ensure efficient optical coupling. The discussed hybrid quantum photonics is based on color centers in NDs, in particular the SiV center, interfaced with $\mathrm{Si}_3\mathrm{N}_4$-photonics. We elaborate on the hybrid device development on the following points:
\begin{enumerate}
\item 
The design, simulation and fabrication of photonics based on $\mathrm{Si}_3\mathrm{N}_4$-technology.
\item 
The control of SiV centers in NDs and their integration into $\mathrm{Si}_3\mathrm{N}_4$-photonics.
\item 
The performance of the final hybrid quantum device.
\end{enumerate}
The investigated platform holds great potential for future quantum technology due to the following facts:
\begin{enumerate}
\item 
The platform profits from all advantages of a well-established photonics platform with very low propagation loss thereby enabling integration into large-scale, highly-functional IPCs.
\item 
The quantum properties of color center in NDs, here in particular the SiV center, can be optimized separately and enables to improve the optical- and spin-properties even beyond to what is possible in bulk diamond.
\item 
The possibility of complete on chip control enables a user-friendly, high bandwidth platform on a small footprint with advantages for scalability. 
\item 
The diamond host comprises long-lived nuclear spins that can potentially be used as quantum storage unit in applications that rely on quantum memories.
\end{enumerate} 

\section{Nanophotonic Device Design of $\mathrm{Si}_3\mathrm{N}_4$-Photonics towards Optimal Hybrid Integration}

This section recapitulates the essential elements of IPCs optimized for its use in hybrid quantum photonics.

\subsection{Single-Mode Planar Optical Waveguide}
\begin{figure}
\includegraphics[width=\textwidth]{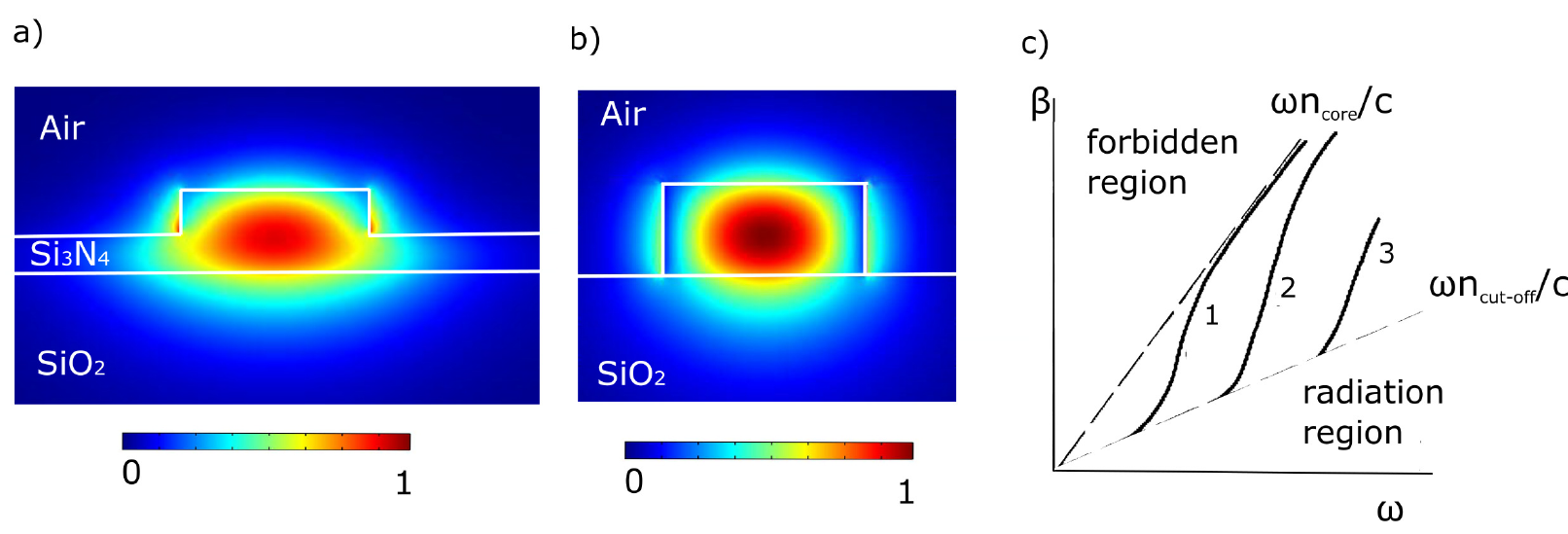}
\caption[]{Simulated electric field profile of the transverse electric fundamental (TE0) mode on $\lambda=738\nm$ in a single mode (a) half-etched rib waveguide (width $450\nm$, height of rib $100\nm$) and (b) fully-etched ridge waveguide (width $450\nm$, height $200\nm$). (c) $\beta-\omega$ dispersion diagram for a dielectric planar waveguide.\\ 
Copyright declaration: First published in the PhD Thesis of Dr. Anna P. Ovvyan \lq{}\lq{}Nanophotonic circuits for single photon emitters" DOI: 10.5445/IR/1000093929}
\label{fig:single_mode_waveguide}
\end{figure}

The key element of all IPCs is the planar waveguide. It confines and guides light within the photonic circuit and ensures the interconnection between the photonic devices on the chip. The planar waveguide consists of a core with high refractive index surrounded by a cladding material of lower refractive index, thus enabling to confine light to propagate in one direction (Fig. \ref{fig:single_mode_waveguide}).
The $\mathrm{Si}_3\mathrm{N}_4$-platform is one of the most prominent platforms for the realization of integrated photonic circuits on-chip. It provides sufficiently high refractive index contrast from $\mathrm{Si}_3\mathrm{N}_4$ to $\mathrm{SiO}_2$ enabling tight mode confinement in the wavequide (Fig. \ref{fig:single_mode_waveguide} a, b) allowing for photonic circuits with a small footprints. Furthermore, $\mathrm{Si}_3\mathrm{N}_4$ provides low-loss broadband optical transparency covering the entire visible range up to the near infrared spectrum.
The guided modes in planar waveguide are the solutions of Maxwell’s equations for particular boundary conditions of the waveguide and materials expressed as moving planewaves \cite{lefebvrePhotoluminescenceImagingSuspended2006}:

\begin{equation}
E_m = E_{0m}(y,z)e^{\I\omega t - \beta_m x}
\end{equation}
\begin{equation}
H_m = H_{0m}(y,z)e^{\I\omega t - \beta_m x},
\end{equation}
where $E_{0m}(y,z)$ and $H_{0m}(y,z)$ is the profile of the electric and magnetic fields, $\beta_m = \frac{2 \pi}{\lambda}n_{\mathrm{eff},m}$ is the propagation constant and m being the mode index. The modes are discrete, since they are trapped in the finite volume of the core of the waveguide. Transverse electric (TE)-modes are the guided modes of the planar waveguide, where the only non-vanishing fields are $E_y\neq0, H_x\neq0, H_z\neq0$. The transverse magnetic (TM)-modes have the non-vanishing field components $H_y\neq0,E_x\neq0, E_z\neq0$. Therefore, there is only one non-vanishing component in longitudinal direction for each mode of the planar waveguide ($H_x$ for TE-mode and $E_x$ for TM mode). It should be noted, that transverse electric magnetic modes (TEM) ($H_x=0$ and $E_x=0$) and hybrid modes ($E_y\neq0$ and $H_y\neq0$) do not exist in planar waveguides \cite{lefebvrePhotoluminescenceImagingSuspended2006}.
A typical $\beta-\omega$ dispersion diagram is illustrated in Fig.\ref{fig:single_mode_waveguide} c) for three discrete guided modes. The radiation continuous modes are mostly propagating in the cladding as not-guided modes. The propagation constant and the effective refractive index of the guided modes should satisfy the following inequalities
\begin{equation}
\frac{\omega}{c} n_\mathrm{cut-off} < \beta_m < \frac{\omega}{c} n_\mathrm{core}\;,
\end{equation}
\begin{equation}
n_\mathrm{cut-off} < n_\mathrm{eff,m} < n_\mathrm{core}\;.
\end{equation}

\subsection{Apodized Grating Coupler}
\begin{figure}
\includegraphics[width=\textwidth]{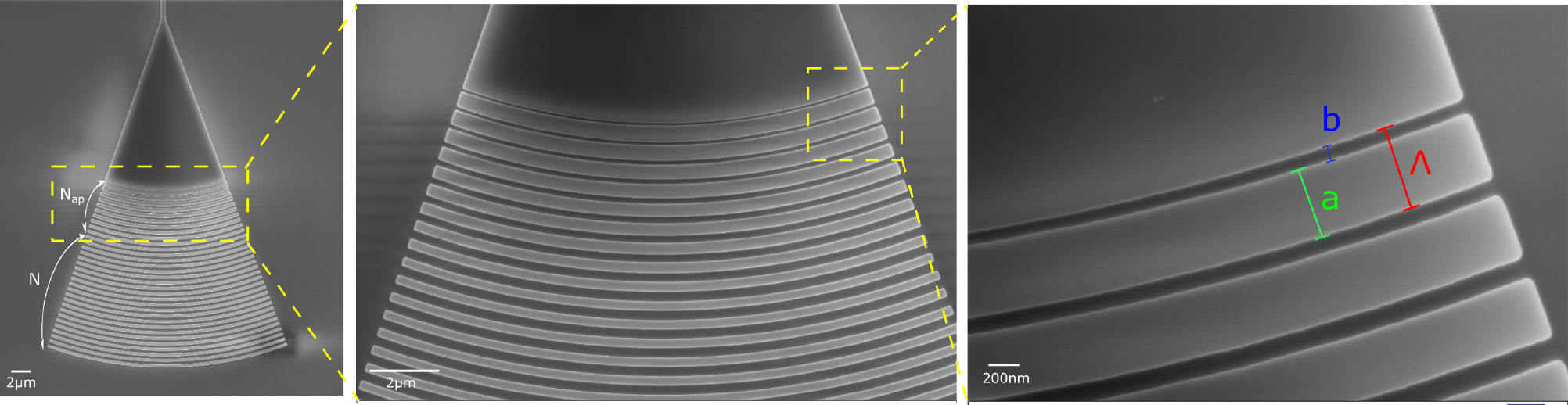}
\caption[]{Scanning electron microscope (SEM) image of an apodized grating coupler with a period $\Lambda$ (\textit{left panel}) and zoom-ins in (\textit{middle panel}) and (\textit{right panel}).\\ 
	Copyright declaration: First published in the PhD Thesis of Dr. Anna P. Ovvyan \lq{}\lq{}Nanophotonic circuits for single photon emitters" DOI: 10.5445/IR/1000093929}
\label{fig:sem_apodized_grating_coupler}
\end{figure}

\begin{figure}
\includegraphics[width=\textwidth]{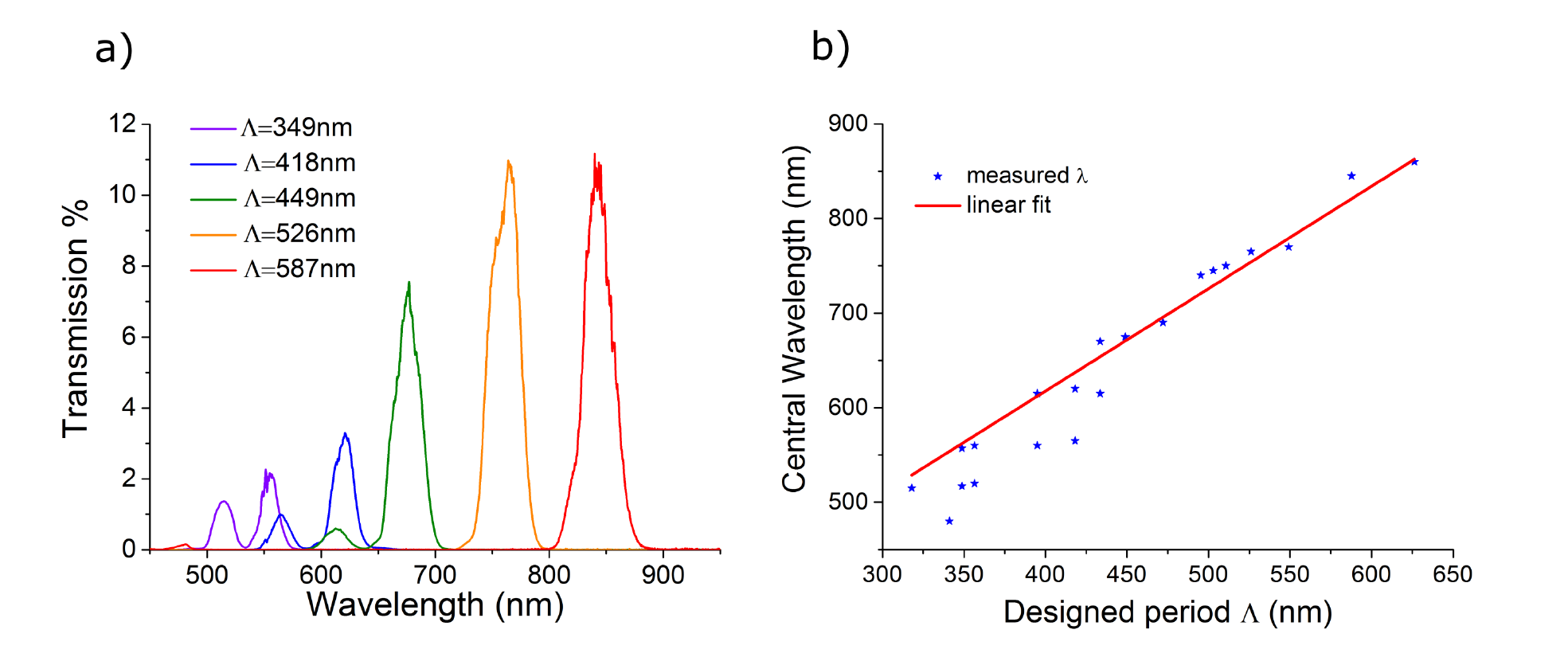}
\caption[]{(a) Measured Normalized Transmission spectrum of a fabricated half-etched waveguide terminated by two grating couplers with variation of their periods $\Lambda$. (b) Measured central wavelength as a function of designed period $\Lambda$ of the coupler.\\ 
	Copyright declaration: First published in the PhD Thesis of Dr. Anna P. Ovvyan \lq{}\lq{}Nanophotonic circuits for single photon emitters" DOI: 10.5445/IR/1000093929}
\label{fig:transmission_apodized_grating_coupler}
\end{figure}

Integrated on-chip photonic elements are interconnected with optical waveguides. In order to couple light in and out of the chip grating couplers are used which convert incident Gaussian-like, free-space modes or modes from an optical fiber into the waveguide mode and vice versa. The periodic structure of the coupler, shown in Fig. \ref{fig:sem_apodized_grating_coupler}, provides out of plane coupling via diffraction from every tooth of the device due to the Bragg condition \cite{braggReflectionXraysCrystals1913}
\begin{equation}
\sin\left(\theta_{in}\right) = \frac{n_{\mathrm{eff}} - m \lambda /\Lambda}{n_{\mathrm{clad}}}\;,
\end{equation}
where $n_{\mathrm{eff}}$ is the effective refractive index of the mode in the grating, $n_{\mathrm{clad}}$ is the refractive index of the top cladding of the grating, $m$ is the diffraction number, $\Lambda$ is the period of the grating and $\lambda$ is the wavelength of the input light.
The coupling efficiency is improved by using the recipe of apodization with a filling factor $ff=a/\Lambda$. Also, an optimization of the length of the grating coupler's taper ensures the matching of the fiber-coupled, input mode of a Gaussian-like free-space mode \cite{bozzolaOptimisingApodizedGrating2015,romero-garciaSiliconNitrideCMOScompatible2013,chenApodizedWaveguideGrating2010} and the waveguide mode. Variation of the period $a$ of the grating is the tuning mechanism for the diffracted central wavelength. The measured transmission spectrum of the photonic device, shown in Fig. \ref{fig:transmission_apodized_grating_coupler}, is recorded for a waveguide terminated by two apodized grating couplers, as shown in Fig. \ref{fig:sem_apodized_grating_coupler} a). 
The grating couplers act as spectral and polarizing filter, which provide an alignment-free approach to couple light in and out of the photonic circuit by converting the incident Gaussian-like free space mode into a guided mode \cite{lomonteEfficientSelfimagingGrating2021}. Furthermore, they are fabricated in the same electron-beam lithography step as the photonic circuit on-chip and ensure a compact footprint and easy access to any device on the chip.

\subsection{Mach-Zehnder Interferometer}
\begin{figure}
\includegraphics[width=\textwidth]{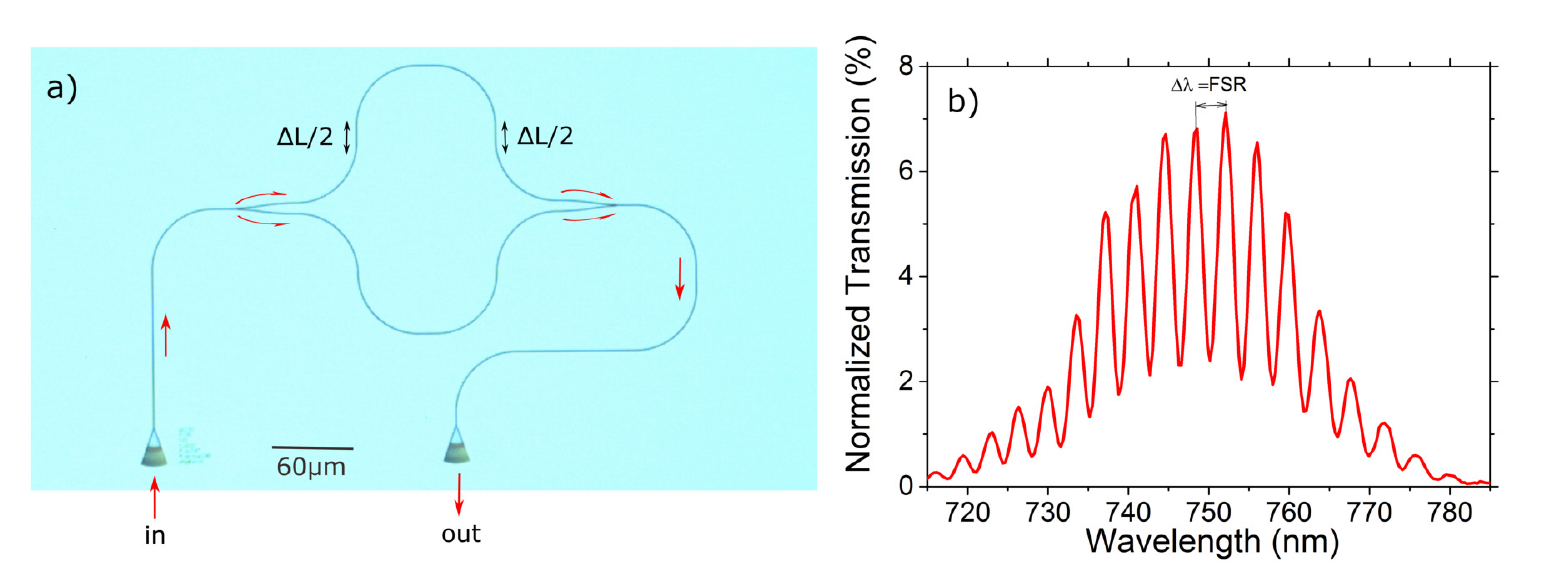}
\caption[]{Mach-Zehnder interferometer on-chip. (a) Optical micrograph of a fabricated MZI on $\mathrm{Si}_3\mathrm{N}_4$ chip with path difference $\Delta l$. (b) Normalized transmission spectrum of a MZI measured with a supercontinuum white light source, where the outcoupled light is detected with a spectrometer. The extinction ratio (ER) at $\lambda=750\nm$ equals $\mathrm{ER}\approx10\cdot(\log{\left(0.07/0.02\right))}=5.4\,\mathrm{dB}$. This low value is due to a non-ideal 50/50 splitting ratio as a result of non-optimized splitter and combiner parameters, and because of the low temporal coherence of the supercontinuum light, i.e. short coherence length.\\ 
	Copyright declaration: First published in the PhD Thesis of Dr. Anna P. Ovvyan \lq{}\lq{}Nanophotonic circuits for single photon emitters" DOI: 10.5445/IR/1000093929}
\label{fig:on_chip_MZI}
\end{figure}
The Mach-Zehnder interferometer (MZI) is the fundamental element of amplitude and phase modulators on chip. In particular, thermo-optical (TO) tunable filters are used, where a thermally induced phase difference results in the shift of the interference fringes in the frequency space. In a MZI on chip, as depicted in Fig. \ref{fig:on_chip_MZI} a), the input light is equally split by a Y-splitter into two beams propagating in two waveguide-arms. The waveguide-arms have a certain length difference and are then recombined on a Y-combiner. The resulting interference pattern is shown in Fig. \ref{fig:on_chip_MZI} b). The fixed path difference $\Delta l$ between both arms introduces the corresponding phase difference of $\Delta\varphi(\lambda)=\frac{2\pi}{\lambda}\cdot\Delta l \cdot n_{\mathrm{eff}}$, where $n_{\mathrm{eff}}$ is the effective refractive index of the waveguide mode. Accordingly, the intensity at the output port is modulated as
\begin{eqnarray}
I_{1,\mathrm{MZI}}\left(\lambda\right)&=&\frac{1}{4}\cdot I_0\cdot\left(e^{-\I\left(\varphi_1+\Delta\varphi\right)}\cdot e^{-\frac{\alpha}{2}\left(l+\Delta l\right)}+e^{-\I\varphi_1}\cdot e^{-\frac{\alpha}{2}l}\right)\cdot\left(\mathrm{h.c.}\right) \nonumber\\
&=&\frac{1}{4}\cdot\ I_0\cdot\left(2\cdot \cos{\left(\mathrm{\Delta\varphi}\right)}\cdot e^{-\alpha\left(l+\frac{\mathrm{\Delta l}}{2}\right)}+e^{-\alpha l}+e^{-\alpha\left(l+\mathrm{\Delta l}\right)}\right),
\label{eq:IMZI}                     
\end{eqnarray}
where $\varphi_1$ is the  phase of the mode in the shortest arm, $\Delta l$ is the length difference between both arms and $e^{-\frac{\alpha}{2}\left(l+\Delta l\right)}$ and $e^{-\ \frac{\alpha}{2}l}$ accounts for the propagation loss in the arms respectively.
In case of negligible loss, the intensity at the output port becomes
\begin{equation}
I_{1,\mathrm{MZI}}\left(\lambda\right)=\frac{1}{2}\cdot\left(1+\cos{\left(\Delta\varphi\left(\lambda\right)\right)}\right)\;.
\end{equation}
The free spectral range is determined by
\begin{equation}
\Delta\lambda=\frac{\lambda^2}{n_{\mathrm{gr}}\cdot\Delta l}\;,
\end{equation}
where $n_{\mathrm{gr}}$ is the group refractive index of the waveguide mode. The extinction ratio (ER) is the contrast of the interference fringes and strongly depends on the optical losses and the splitting ratio of the Y-splitter (ideally 50/50). It can be derived from (\ref{eq:IMZI}) as
\begin{equation}
\mathrm{ER}=\frac{I_{\mathrm{max}}}{I_{\mathrm{min}}}=\left(\frac{1+e^{-\frac{\alpha}{2}\Delta l}}{1-e^{-\frac{\alpha}{2}\Delta l}}\right)^2\;.
\end{equation}
MZIs are widely used as on chip ultra-fast modulators \cite{thomsonHighPerformanceMach2013,samaniSiliconPhotonicMach2019,lomonteSinglephotonDetectionCryogenic2021,chenApodizedWaveguideGrating2010} and biosensors.

\subsection{Mach-Zehnder Interferometer Thermo-Optical Tunable Filter}

The realization of all-integrated, scalable photonic circuits with accessible quantum emitters requires optical filtration of the excitation light with a high extinction ratio to distinguish the emitted single photons from the background signal. The aim is to realize a compact and tunable optical filter which can be integrated in nanophotonic circuits on-chip. Cascading several filter stages provides an enhancement of the overall filtration depth. The tunability of the filtering wavelength could be provided by tuning the refractive index via the thermo-optical effect \cite{densmoreCompactLowPower2009,nedeljkovicMidInfraredThermoOpticModulators2014,chuCompactThermoopticSwitches2005,espinolaFastLowpowerThermooptic2003,harrisEfficientCompactLow2014,wangSOIThermoopticModulator2003} which can be controlled with local micro-heaters integrated on chip. Such filters can be integrated in photonic circuits with an evanescently coupled single photon emitter, such as the SiV-center.
\begin{figure}
	\includegraphics[width=\textwidth]{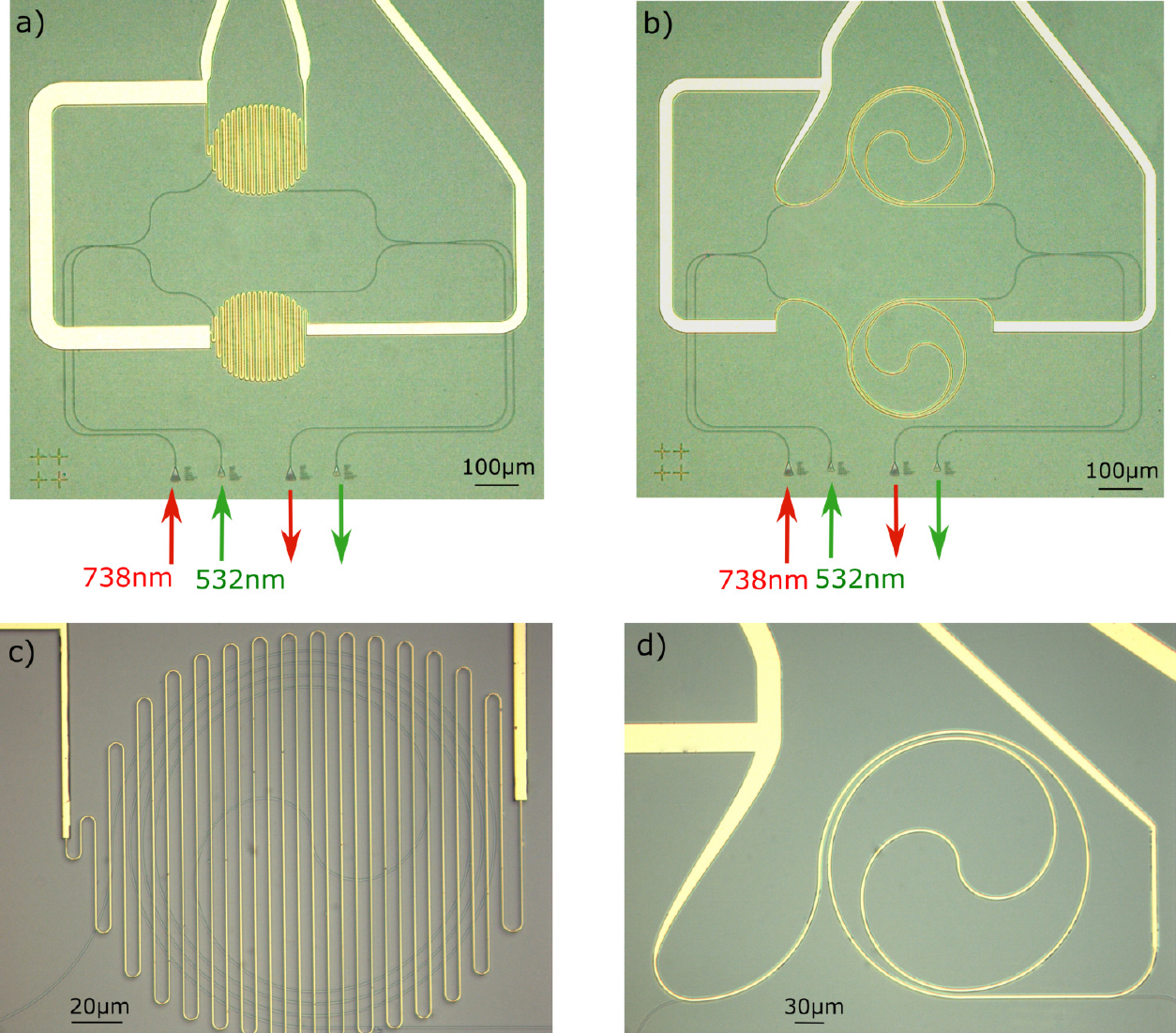}
	\caption[]{Optical micrographs of thermo-optical single MZIs equipped with (a) double-meander and (b) double-spiral shaped microheaters on top of both arms of the MZI with zoom-ins on (c) meander ($\mathrm{width}=0.5\,\umu\mathrm{m}$) and on (d) spiral-shaped ($\mathrm{width}=3\,\umu\mathrm{m}$) microheaters, each with a spiral-folded $\mathrm{Si}_3\mathrm{N}_4$ waveguide beneath.\\ 
	Copyright declaration: First published in the PhD Thesis of Dr. Anna P. Ovvyan \lq{}\lq{}Nanophotonic circuits for single photon emitters" DOI: 10.5445/IR/1000093929}
	\label{fig:Si3N4_MZI}
\end{figure}\\
The here discussed single TO tunable MZIs are equipped with spiral and meander shaped microheaters. The design enables to simultaneously obtain maximum and minimum transmission on the desired wavelengths, in particular the wavelength of the SiV center fluorescence, at $\lambda=740\nm$, and the excitation light, at $\lambda=532\nm$, respectively. MZI photonic devices on-chip consist of two branches, namely long waveguides which are recombined on a Y-splitter. A dynamic change of the optical path difference between both arms leads to amplitude tunability of the mode at the necessary wavelength of the MZI filter. The dynamic change can be obtained either by altering the effective refractive index of the propagating mode in one of the arms or by changing the physical length difference $\Delta l$ between both arms, whereas the last is fixed in the nanophotonic circuits on-chip. The result is a thermally induced phase shift $\Delta \varphi$ for the light travelling through one of the arms of the MZI. The waveguide in one of the branches should experience a temperature change $\Delta T$ which leads to a change of the effective refractive index of the propagating mode due to the TO effect. This can be realized by placing a metal microheaters on top of the MZI branch. To avoid optical absorption loss due to the presence of the metal in the vicinity of the waveguide, the nanophotonic structures should be covered with a cladding layer to isolate the propagating optical mode from the atop metal microheaters. The TO effect provides a rather weak modification of the refractive index due to the small thermo-optical coefficient of silicon nitride, $\frac{dn}{dT}\approx{10}^{-5} K^{-1}$ \cite{tuThermalIndependentSiliconNitride2012}, which is approximately ten times lower than in silicon. Therefore, the branches of the MZI should consist of long waveguides, several $\mathrm{mm}$, in order to significantly change the phase difference during the heating and thus increase the shift of the interference fringes of the MZI.
The waveguides are folded in spiral design to concentrate the heat in the center and thus to increase the efficiency of the heating as well as to provide a compact device footprint. The developed MZI filters are composed of two Y-splitters and two long waveguides folded in a spiral pattern, featuring a small path difference. Two types of devices are realized where the arms of the MZI waveguides are equipped with spiral or meander design of microheaters as shown in Fig. \ref{fig:Si3N4_MZI}. The metal heater wire is aligned with the center of the waveguide to maximize the overlap of the heat concentration with the guided optical mode.\\
When applying a voltage to the microheater, the area beneath the waveguide is locally heated, inducing a phase shift $\varphi_{\mathrm{shift}}$ of the propagating mode in the MZI arm. The phase difference at the output of the thermo-optical MZI modulator after heating the longer arm can be expressed as the following:
\begin{equation}
	\Delta\varphi=\Delta\varphi_{\mathrm{fixed}}+\varphi_{\mathrm{shift}}=\frac{2\pi}{\lambda}\Delta l\cdot n_{\mathrm{eff}}+\frac{2\pi}{\lambda}(l_{\mathrm{spiral}}+\Delta l)\cdot \frac{\D n}{\D T}\Delta T\;.
	\label{eq:MZI_phase_long}
\end{equation}
Heating of the shorter arm yields the phase difference
\begin{equation}
	\Delta\varphi=\Delta\varphi_{\mathrm{fixed}}-\varphi_{\mathrm{shift}}=\frac{2\pi}{\lambda}\Delta l\cdot n_{\mathrm{eff}}-\frac{2\pi}{\lambda}(l_{\mathrm{spiral}})\cdot \frac{\D n}{\D T}\Delta T\;,\label{eq:MZI_phase_short}
\end{equation}
where $\lambda$ is the desired wavelength in vacuum, $\Delta l$ is the original fixed length difference between the arms, $l_{\mathrm{spiral}}$ is the length of the spiral-arm waveguide, $n_{\mathrm{eff}}$  is the effective refractive index of the guided mode,  $\frac{\D n}{\D T}$ is the thermo-optical coefficient of silicon nitride and $\Delta T$ is the induced temperature change.
The induced temperature difference, which corresponds to the switching power $P_\pi$, which is necessary in order to induce a $\pi$-phase shift, $\varphi_{\mathrm{shift}}=\pi$, (\ref{eq:MZI_phase_long}, \ref{eq:MZI_phase_short}) is determined as
\begin{equation}
	\Delta T_\pi=\frac{\lambda}{2\cdot l_{\mathrm{spiral}}\cdot \frac{\D n}{\D T}}\;.
\end{equation}
The demonstrated thermo-optical MZI filter, which are equipped with microheaters on top of both arms, make it possible to double the amplitude of the possible shift of the interference fringes. The design enables to apply the required voltage to the microheater on top of both arms separately and independently. Thus, the interference fringes will be either shifted towards shorter or longer wavelengths.\\
An important characteristics of the designed device is the optical propagation loss, which is determined by measuring the transmission of the photonic devices containing spiral waveguides with variation in length in the range $1$--$6.3\mm$ and covered with a cladding layer on top equipped with microheaters. The measured average propagation loss amounts to $0.49\,\mathrm{dB}/\mathrm{mm}$ at $\lambda=766\nm$ \cite{ovvyanCascadedMachZehnder2016}.\\
\begin{figure}
	\includegraphics[width=\textwidth]{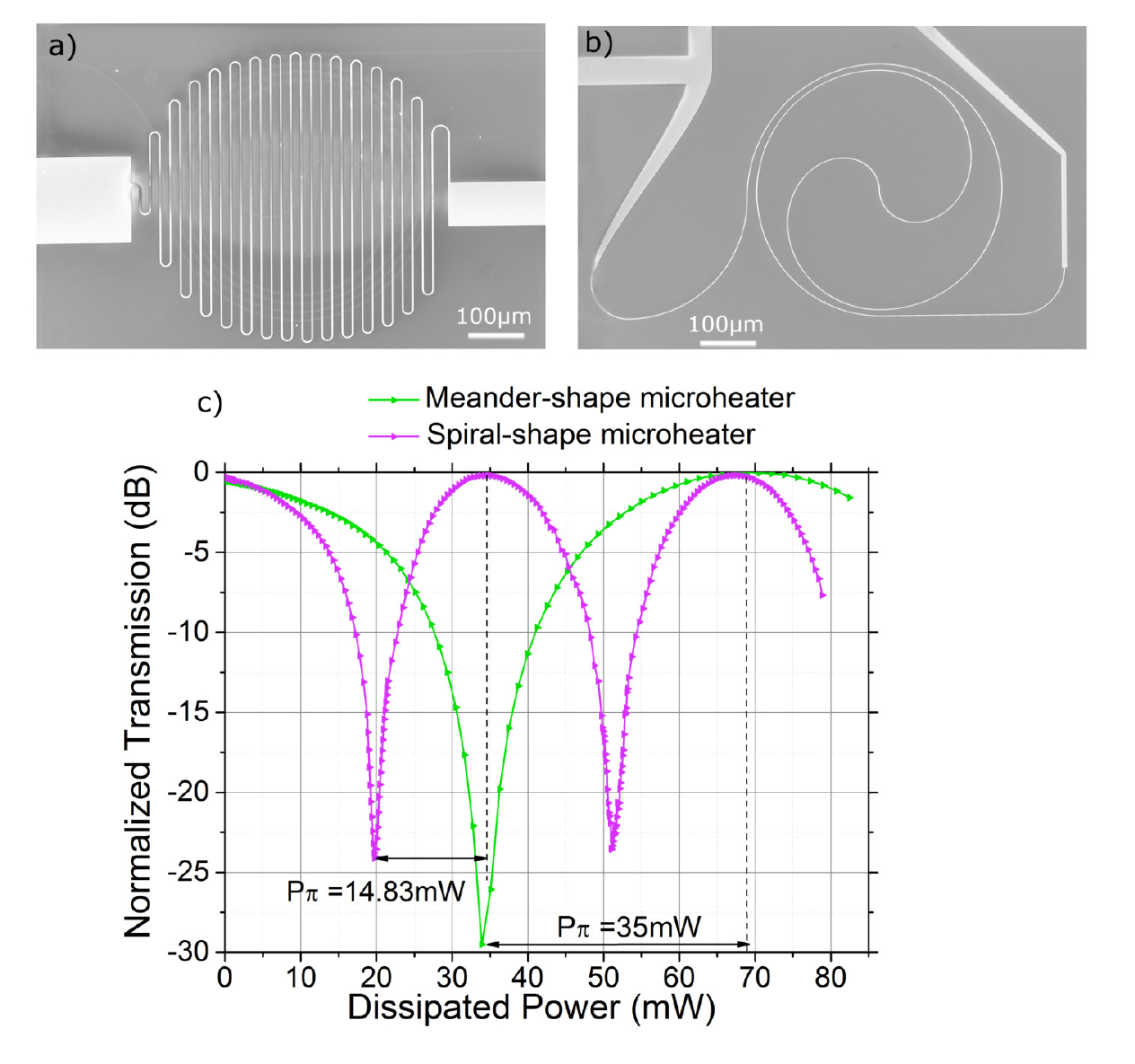}
	\caption[]{SEM images of (a) meander ($\mathrm{width}=800\nm$) and (b) spiral ($\mathrm{width}=800\nm$) architecture microheaters. (c) Modulated normalized transmission of a single MZI device equipped with double-spiral shaped (\textit{violet curve}) and double-meander shaped (\textit{green curve}) heaters with identical $800\nm$ heater width. The voltage is applied to the top microheater of each MZI device.
The difference in peak positions are explained by the different switching powers of the spiral- and meander-shaped microheater MZIs.\\
		Copyright declaration: First published in the PhD Thesis of Dr. Anna P. Ovvyan \lq{}\lq{}Nanophotonic circuits for single photon emitters" DOI: 10.5445/IR/1000093929}
	\label{fig:Transmission_vs_disip_power_MZI}
\end{figure}
The aim is to optimize the configuration for a filter with minimized switching power $P_\pi$ to obtain a phase shift $\varphi_{\mathrm{shift}}=\pi$. A typical modulated normalized transmission spectrum for the thermo-optical MZI devices with spiral-(violet curve) and meander-geometry (green curve) of the microheaters is shown in Fig. \ref{fig:Transmission_vs_disip_power_MZI}. For both measurements, the heater width is $800\,\nm$. Only one arm is heated, while the optical input wavelength is kept fixed at $\lambda=766\nm$. The MZI device with spiral-shaped microheaters shows more than two times lower switching power $P_\pi=14.83\mW$ in comparison to the meander-shaped microheater at $P_\pi=35\mW$. 
By optimizing the geometry of the microheaters and increasing the active length of the waveguide, the switching power can be further reduced.
The most efficient thermo-optical MZI device achieves a switching power of $P_\pi=12.2\mW$ with spiral-shaped heaters with a width of $200\nm$ \cite{ovvyanCascadedMachZehnder2016}. This design achieves an ER of $20\,\mathrm{dB}$. When working at $\lambda=738\nm$ the induced temperature change for the considered MZI is $\Delta T=20\,\mathrm{K}$. 

\subsection{Cascaded Mach-Zehnder Interferometer as Tunable Filter on Chip}
\begin{figure}
\includegraphics[width=\textwidth]{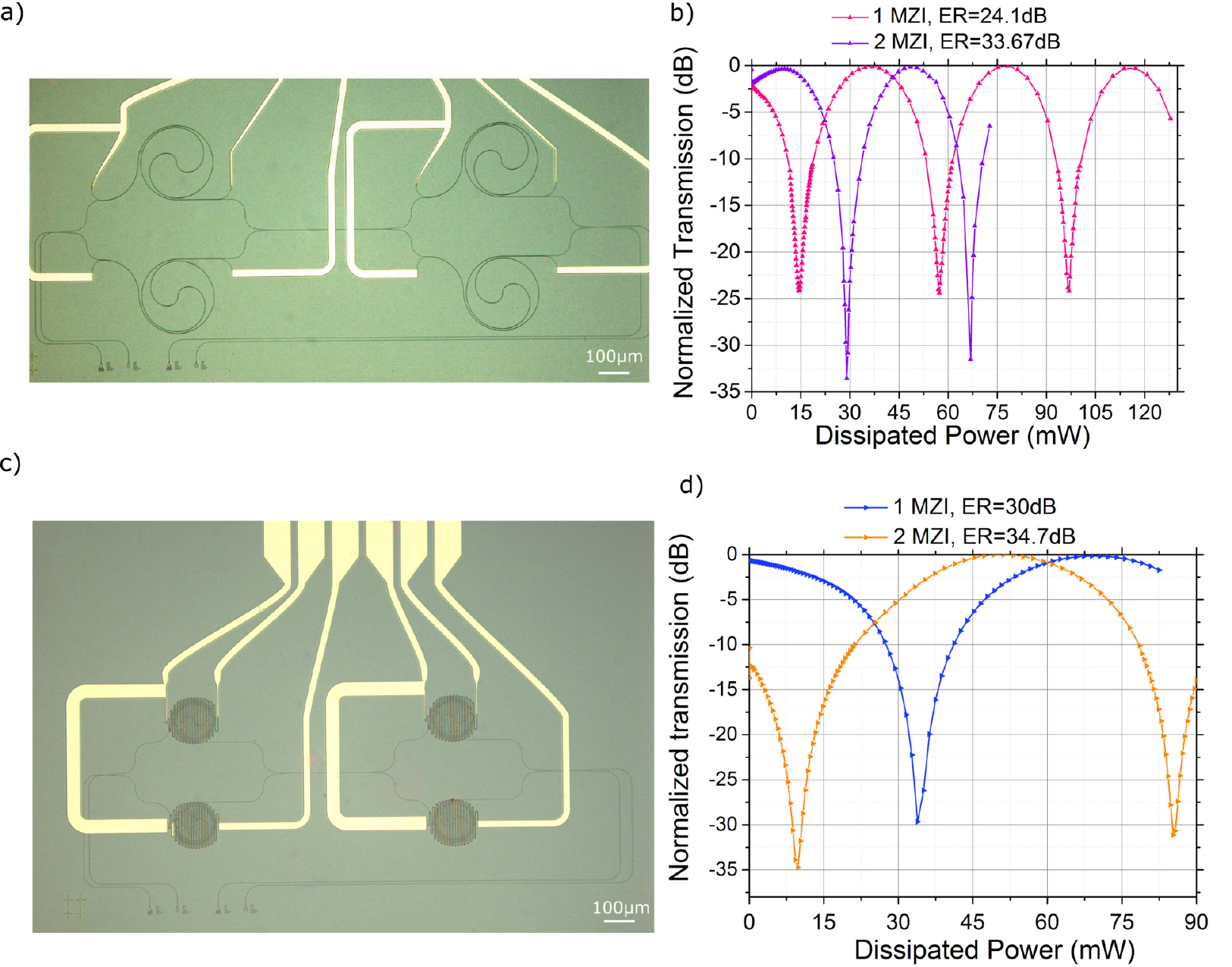}
\caption[]{Optical micrographs of cascaded thermo-optical MZI devices equipped with (a) double-spiral shaped microheater and (c) double-meander shape. Comparison of the normalized transmission through a single thermo-optical MZI device and a cascade of two MZIs supplied with a (b) double-spiral shaped microheater and (d) double-meander shaped. The length of the spiral waveguide in both MZIs is $1.78\mm$.\\ 
	Copyright declaration: Firstly published in the PhD Thesis of Dr. Anna P. Ovvyan "Nanophotonic circuits for single photon emitters"}
\label{fig:Cascaded_MZI}
\end{figure}

Cascaded TO devices, as shown in Fig. \ref{fig:Cascaded_MZI}, can be employed to further improve the filtration depth. The maximum ER of such a filter can be obtained when the interference patterns of both MZIs in cascade are perfectly overlapped. This can be achieved by applying a scanning voltage to one microheater of one of the cascaded MZIs to thermally shift and align the interference pattern of this MZI with respect to the interference pattern of the second MZI.\\
The modulated normalized transmission of a single thermo-optical MZI and of a cascaded thermo-optical MZIs equipped with spiral- (a, b) and meander- (c, d) design microheaters are shown in Fig. \ref{fig:Cascaded_MZI}. An increase of the ER by $9.6\,\mathrm{dB}$ and by $4.7\,\mathrm{dB}$ is demonstrated for devices with a spiral-heater and meander-heater, respectively. The cascaded MZI devices equipped with spiral microheaters are more efficient in terms of power consumption as compared to meander microheaters. However, the extinction ratio of both types of devices are similar and independent of the shape since in both cases the buffer layer covers the nanophotonic devices and prevents from absorption of the light.\\
By simultaneously applying two independent scanning electrical voltages to one of the microheaters of the first MZI (\lq{}external’ loop) and one of the microheaters of the second MZI (\lq{}internal' loop) a corresponding phase shift in each MZI is induced. Respectively, the interference patterns are overlapped depending on the induced phase shift. Thus, the modulated intensity at the output of two cascaded MZIs (assuming 50/50 splitting and without accounting for propagation loss) is determined as 
\begin{eqnarray}
I_{2,\mathrm{MZI}} &=& \frac{1}{16}\cdot(4 + 4\cos\left(\Delta \varphi_1 \right)+4\cos\left(\Delta \varphi_2 \right) \nonumber \\ 
&& + 2\cos\left(\Delta \varphi_1 + \Delta \varphi_2 \right)+2\cos\left(\Delta \varphi_1 - \Delta \varphi_2 \right))\;,
\label{eq:two_MZI}
\end{eqnarray}
where $\Delta \varphi_1$ is the phase difference in the first MZI and $\Delta \varphi_2$ is the phase difference in the second MZI. $\Delta \varphi_1$ and $\Delta \varphi_2$ can be found by (\ref{eq:MZI_phase_long}, \ref{eq:MZI_phase_short}) and depend on which arm was heated.\\
\begin{figure}
\includegraphics[width=\textwidth]{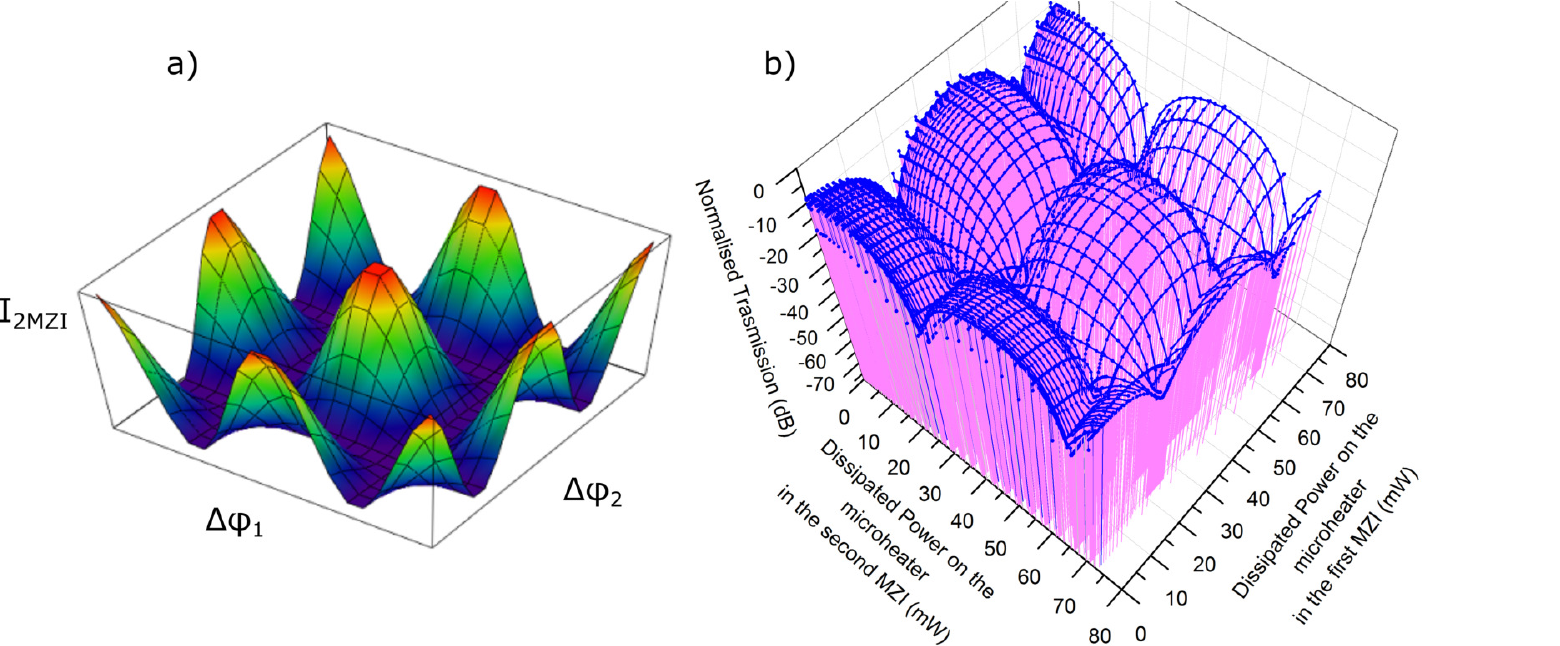}
\caption[]{Analytical modulated intensity at the output of two MZIs as a function of phase difference $\Delta \varphi_1$ and $\Delta \varphi_2$ of the first and second MZI, respectively, according to (\ref{eq:two_MZI}). b) Measured normalized transmission of two cascaded spiral-arm MZIs equipped with spiral-shape microheaters ($\mathrm{width}=1\,\umu\mathrm{m}$) by employing a voltage on top one of the arms of first and second MZIs.\\ 
	Copyright declaration: Firstly published in the PhD Thesis of Dr. Anna P. Ovvyan "Nanophotonic circuits for single photon emitters"}
\label{fig:Analytical_MZI}
\end{figure}
The analytically determined modulated transmission is shown in Fig. \ref{fig:Analytical_MZI} a), whereas the measured normalized transmission at the output of the two thermo-optical cascaded MZI devices is displayed in b), while two scanning voltages were applied to one of the microheaters of the first MZI and one of the microheaters of the second MZI simultaneously. The theoretical and experimental results are in a good agreement, taking into account that the dissipated power in the microheater is proportional to $\Delta \varphi$.

\subsection*{Tuning Performance of Single and Cascaded Thermo-Optical MZI Filters}
\begin{figure}
\includegraphics[width=0.6\textwidth]{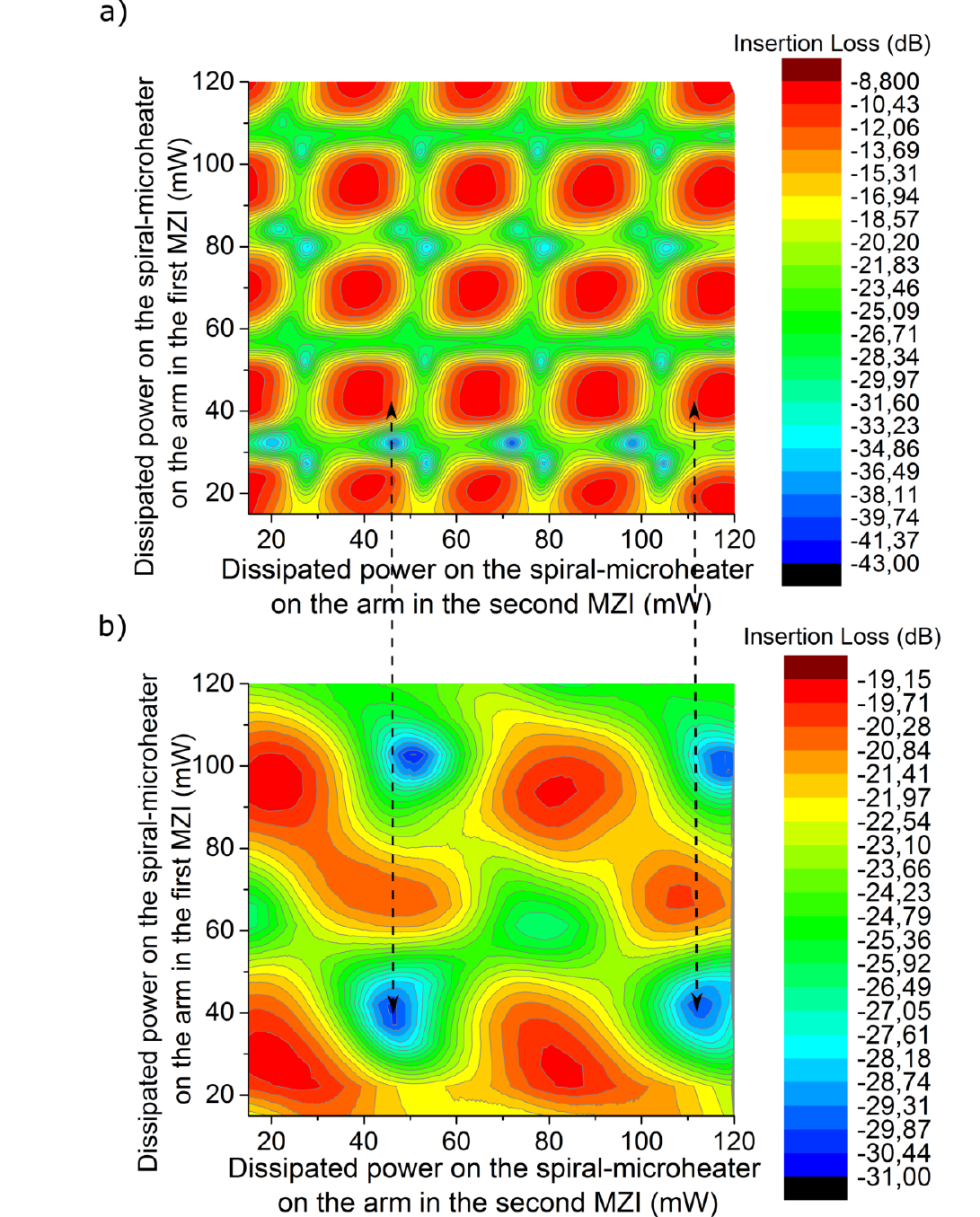}
\caption[]{Transmission of two cascaded thermo-optical MZIs equipped with spiral-shaped microheaters as a function of dissipated power by employing a voltage to the microheaters on top of one of the arms of the first and second MZIs. a) The contour graph shows transmission at $738\nm$, $\mathrm{ER}=32.9\,\mathrm{dB}$. b) Contour graph displays transmission at $532\nm$, $\mathrm{ER}=11.35\,\mathrm{dB}$. Parameters of MZI: width of heater $200\nm$, minimum radius of spiral-folded waveguide $130\,\umu\mathrm{m}$.\\ 
	Copyright declaration: Firstly published in the PhD Thesis of Dr. Anna P. Ovvyan "Nanophotonic circuits for single photon emitters"}
\label{fig:transmission_two_cascaded_MZI}
\end{figure}

The developed MZI thermo-optical tunable filters can be integrated in hybrid nanophotonic circuits on-chip with coupled single photon emitters. The excitation light is blocked while the emitted fluorescence light is transmitted, simultaneously.
\begin{figure}
\includegraphics[width=0.6\textwidth]{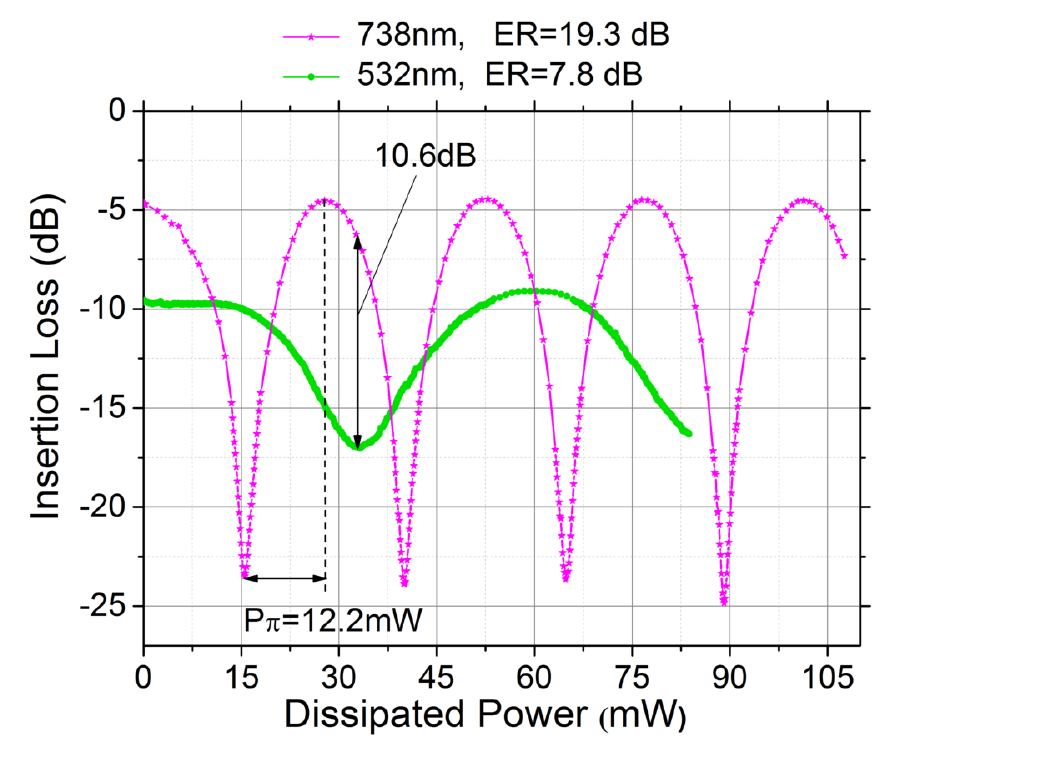}
\caption[]{Modulated insertion loss a thermo-optical spiral-folded arms single MZI equipped with double-spiral architecture microheater with inner radius of bend $130\,\umu\mathrm{m}$ and $\mathrm{length}=1.78\mm$. Pink curve corresponds to measured transmission at $738\nm$, green curve corresponds to measurement performed at $532\nm$.\\ 
	Copyright declaration: Firstly published in the PhD Thesis of Dr. Anna P. Ovvyan "Nanophotonic circuits for single photon emitters"}
\label{fig:modulated_insertion_MZI}
\end{figure}
In particular, the thermo-optical single and cascaded MZIs shown in Fig. \ref{fig:Si3N4_MZI} and \ref{fig:Cascaded_MZI} are intended to be utilized as a tunable filter in the nanophotonic circuit with coupled SiV color centers. Therefore, the filters should enable to transmit light at a wavelength around $\lambda=738\nm$ while cutting the excitation light at a wavelength of about $532\nm$. The measurements are performed by delivering both laser lights to the thermo-optical MZI.\\
The performance of a single thermo-optical MZI filter with an optimized architecture (narrow spiral-shape microheater with $200\nm$ width) is shown in Fig. \ref{fig:modulated_insertion_MZI}. By applying $P = 33\,\mathrm{mW}$ to the long arm of a single MZI yields a $10.6\,\mathrm{dB}$ filtration characteristic. The transmitted signal at a wavelength of $\lambda=532\nm$ (green curve in Fig. \ref{fig:modulated_insertion_MZI}) reduces to $-17\,\mathrm{dB}$, whereas the transmission at a wavelength of $738\nm$ (red curve in Fig. \ref{fig:modulated_insertion_MZI}) yields $-6.4\,\mathrm{dB}$. The experimentally measured ER of one MZI is higher when coupling light at $\lambda=738\nm$ ($\mathrm{ER}=19.3\,\mathrm{dB}$) than for coupled light at $532\nm$ (Fig. \ref{fig:modulated_insertion_MZI}) since the waveguide is designed to be single mode for $\lambda=738\nm$. However, this waveguide is not single mode at $\lambda=532\nm$ leading to a degradation of the interference pattern and a decreasing ER.\\
The filtration characteristic is improved when utilization the cascaded thermo-optical MZI as shown in Fig. \ref{fig:Cascaded_MZI} a) equipped with spiral-shaped microheaters. By applying two scanning voltages to one of the microheaters of the first MZI and one of the microheaters of the second MZI shows that the measured transmission of cascaded MZIs is modulated as a function of the dissipated power on the microheater of the first and second MZI while coupling light at $\lambda=738\nm$ and $\lambda=532\nm$ respectively, as shown in Fig. \ref{fig:transmission_two_cascaded_MZI}. The maximum ER of the cascaded thermo-optical MZI device reaches $32.9\,\mathrm{dB}$, $13.6\,\mathrm{dB}$ higher than with a single MZI filter of the same geometry. The ER at $\lambda=532\nm$ reaches $11.35\,\mathrm{dB}$, which is $3.55\,\mathrm{dB}$ higher than in the case of a single MZI while coupling light of the same wavelength.
Thus, the maximum filtration characteristics is obtained at the point of maximum transmission of the light at $738\nm$ and suppression of light at $532\nm$. For the explored cascaded thermo-optical MZI this condition is satisfied when the applied electrical power on the long arm of the second MZI is $43\,\mathrm{mW}$ and the electrical power on the long arm of the first MZI is $40\,\mathrm{mW}$. The obtained filtration depth amounts to $20\,\mathrm{dB}$ which exceeds  that of a single MZI by $10\,\mathrm{dB}$ \cite{ovvyanCascadedMachZehnder2016}. This design could in the future be used to separate the  zero phonon emission line (ZPL) from the excitation laser for on chip quantum optical experiments.

\subsection{Photonic Crystal Cavity}
In 1887 Lord Rayleigh explored that a periodic structure of dielectric materials, which is characterized by a periodic dielectric permittivity, introduces a photonic bandgap \cite{rayleighXVIIMaintenanceVibrations1887, lomonteSinglephotonDetectionCryogenic2021}, i.e. a continuous frequency range within which there are
no propagating solutions of Maxwell's equations for any wave vector. These modes decay exponentially inside the periodic structure along the way of propagation. The reason for this decay is that light is partially reflected at each dielectric layer interface. Therefore, these waves destructively interfere along the propagation way. In contrast, light in the wavelength range outside the photonic bandgap propagates through the crystal without scattering, since the wavevector is conserved. Such a photonic structure is the basis of Bragg mirrors.

\subsection*{Origin of Resonance Modes}
The 1D version of a PhC is well-known as Bragg reflector (mirror) consisting of air holes in a semiconductor material. The introduction of a defect in such a Bragg reflector, for instance by changing the spacing between two central holes, leads to a line-defect cavity which efficiently traps light in the area of the defect forming a localized mode. The resonance modes with an energy within the bandgap can be excited. The modes are discrete, since the modes are localized in a finite region, i.e. between two Bragg mirrors. The modes are confined by the bandgap in the longitudinal direction and by total internal reflection in the two transverse directions. Therefore, the state is localized only in one dimension of the 1D PhC cavity. The superposition of two plane waves traveling in opposite directions, which are here extended states with an exponential decay in the PhC, results in a standing wave. Further, one considers a TE-like standing cavity mode with $E_y$ and $H_z$ as the major field components. The magnetic field distribution of the standing wave is derived as
\begin{equation}
H_z=\cos{\left(\omega t-kx\right)}+\cos{\left(\omega t+kx\right)=2\cdot \cos{\left(\omega t\right)}\cdot \cos{\left(kx\right)}}\;,
\end{equation}
where $\omega$ is the frequency of the resonance mode, $k=\frac{2\pi}{\lambda}n_{eff}$ is the wavevector. Taking  Bragg’s law into account \cite{braggReflectionXraysCrystals1913} $\lambda=2 n_{eff} \cdot a$ and the wavevector $k=\frac{\pi}{a}$ can be calculated with the period $a$ of the PhC.
Magnetic and electric components of the standing wave are shifted by $\frac{\pi}{2}$ in space and time. Therefore, the electric field distribution can be determined as
\begin{eqnarray}
E_y&=&\cos{\left(\omega t+\frac{\pi}{2}-\left(kx+\frac{\pi}{2}\right)\right)}+\cos{\left(\omega t+\frac{\pi}{2}+\left(kx+\frac{\pi}{2}\right)\right)}\nonumber\\&=&2\cdot\sin{\left(\omega t\right)}\cdot\sin{\left(kx\right)}\;.
\end{eqnarray}
The planes of oscillations of $H_z$ and $E_y$ are orthogonal. The Poynting vector is determined as
\begin{equation}
S_x=E_y \cdot H_z=\sin{\left(2\omega t\right)}\cdot\sin{\left(2kx\right)}\;.
\end{equation}
and oscillates with twice the frequency in time and space as compared with the electric and magnetic field distribution of the standing wave.

\subsection*{Quality Factor of the Cavity}

One of the main figures of merit of a PhC cavity is the quality factor $Q$ which characterizes the number of optical round-trips of light inside the cavity until the energy of the resonance mode decays by $\exp\left(-2\pi\right)$. In other words, it indicates the lifetime of the localized mode inside the cavity \cite{joannopoulosPhotonicCrystalsMolding2008}.
There are two main damping mechanisms of the resonance modes for a 1D PhC cavity:
\begin{enumerate}
\item The coupling of the resonance mode into the radiative mode which extends in the air being the decay rate $\frac{1}{Q_{\mathrm{rad}}}$
\item The decay into the dielectric waveguide $\frac{1}{Q_{\mathrm{wg}}}$.
\end{enumerate}
The Quality factor is therefore determined as
\begin{equation}
\frac{1}{Q}=\frac{1}{Q_{\mathrm{wg}}}+\frac{1}{Q_{\mathrm{rad}}}\;.
\end{equation}
Increasing the number of segments in each mirror of the PhC cavity increases $Q_{\mathrm{wg}}$. Thus, ideally the $Q$-value of the 1D PhC cavity with two perfect mirrors of infinite number of periods, saturates and is limited by only $Q_{\mathrm{rad}}$. Therefore, the main aim is to suppress the radiation losses of the resonance mode to increase $Q_{\mathrm{rad}}$.
In order to minimize the radiation losses, i.e. by minimizing the leakage of the resonance modes into the surrounding cladding, here in air, it is necessary that the spatial tails of the Fourier harmonics of the resonance modes stay out of the light cone. This can be obtained by special tapering the envelope function of the electric field of the resonance modes at the cavity edges \cite{akahaneHighQPhotonicNanocavity2003}. A smooth adiabatic transition between the Bloch and the guided modes can be achieved by Gaussian attenuation \cite{akahaneHighQPhotonicNanocavity2003} of the electric field pattern of the resonance mode inside the PhC cavity. This can be achieved by tapering the holes drilled in the waveguide, by tapering the waveguide width or by changing the spacing of the holes \cite{mukherjeeOnestepIntegrationMetal2011,gongNanobeamPhotonicCrystal2010} while keeping the other parameters constant  \cite{quanDeterministicDesignWavelength2011,zhouRefractiveIndexSensing2014,zhangUltrahighQTETM2009,yangHighQHighsensitivityWidthmodulated2015}.  Interestingly, the sinc-function of the electric field envelope enables the complete illumination of the Fourier components from the light cone \cite{mccutcheonDesignSiliconNitride2008}.

\subsection*{Local Density of States and Purcell Enhancement}
The spontaneous emission rate is affected by the surrounding environment. For example, the spontaneous emission rate changes when the emitter is placed in the mode of a cavity, as described by E.M. Purcell and called the Purcell effect \cite{purcellResonanceAbsorptionNuclear1946}. The PhC cavity can enhance or inhibit the spontaneous emission rate of the emitter depending on its position in the cavity field and the detuning of the cavity resonance wavelength, which needs to match the transition wavelength of the emitter, as proved experimentally \cite{huletInhibitedSpontaneousEmission1985}. Fermi’s Golden rule \cite{diracQuantumTheoryEmission1927} states that the spontaneous emission rate is proportional to the density of states (DOS). The DOS is determined as the available number of states for a spontaneously emitted photon at a frequency $\omega^{(n)}$. The DOS is defined by \cite{economouGreenFunctionsQuantum2011,lagendijkResonantMultipleScattering1996}
\begin{equation}
\mathrm{DOS}(\omega)=\sum_{n}{\delta(\omega-\omega^{\left(n\right)}})\;.
\end{equation}
In case when the cavity resonance is much larger than the emitter's optical transition linewidth the cavity resonance can be treated as a continuum while the mode density is much larger than its value in free space \cite{harocheExploringQuantumAtoms2006}. Therefore, the spontaneous emission rate can be enhanced by imposing boundary conditions to the field radiated by the emitter by means of an optical cavity, as long as the cavity mode can be considered as a continuum. The local density of states (LDOS) is introduced to characterize the $\mathrm{DOS}(\omega)$ in a specific position $x$ and in a certain direction $l$, defined by \cite{economouGreenFunctionsQuantum2011,lagendijkResonantMultipleScattering1996}
\begin{equation}\label{eq:LDOS}
{\mathrm{LDOS}}_l(x,\omega)=\sum_{n}{\delta(\omega-\omega^{\left(n\right)}})\cdot\varepsilon(x)\cdot\left|E_l^{\left(n\right)}\left(x\right)\right|^2\;,
\end{equation}
where $\varepsilon\left(x\right)$ is the dielectric permittivity of the material and $E_l^{\left(n\right)}(x)$ is the field component of the standing wave inside the cavity. The $DOS(\omega)$ can be determined as
\begin{equation}
\mathrm{DOS}\left(\omega\right)=\sum_{l=1}^{3}\int{{\mathrm{LDOS}}_l\left(x,\omega\right)\cdot d^3}x\;.
\end{equation}
The radiated power of the dipole source is proportional to the overlap integral between the electric field distribution of the resonance mode localized in the cavity and the electric field of the dipole-source \cite{tafloveAdvancesFDTDComputational2013}. The LDOS is proportional to this radiated power and can be determined as \cite{tafloveAdvancesFDTDComputational2013}
\begin{equation}
{\mathrm{LDOS}}_l\left(x_0,\omega\right)=\frac{4}{\pi}\varepsilon\left(x_0\right)P_l\left(x_0,\omega\right)\;,
\end{equation}
where  $P_l\left(x_0,\omega\right)$ is the power radiated by the linearly polarized source positioned at $x_0$. The resonant cavity mode is approximated with the complex eigenfrequency $\omega=\omega_0-\I\alpha$ , where $\alpha>0$, thus ensuring the exponential decay of the mode in the PhC mirror. The $Q$-factor of the mode can be determined as $Q=\omega_0 / 2\alpha$ \cite{joannopoulosPhotonicCrystalsMolding2008}.
On-resonance LDOS at $\omega^{(n)}$ is derived by \cite{joannopoulosPhotonicCrystalsMolding2008} as
\begin{equation}
{\mathrm{LDOS}}_{\mathrm{res}}\approx\frac{2}{\pi\omega^{\left(n\right)}}\cdot\frac{Q^{\left(n\right)}}{V^{\left(n\right)}}\;,
\end{equation}
where $V^{(n)}=\frac{\int{\varepsilon\left|E^{(n)}\right|^2}}{\max{{\left(\varepsilon\left|E^{(n)}\right|^2\right)}}}$ is referred to as the mode volume \cite{purcellResonanceAbsorptionNuclear1946,wardCalculatingPhotonicGreen1998} and $Q^{(n)}/V^{(n)}$ is proportional to the Purcell factor \cite{purcellResonanceAbsorptionNuclear1946}.\\
The resonance mode of the cavity yields a Lorentzian peak for the total power radiated by the source as well as an LDOS spectrum as shown in \cite{zhangUltrahighQTETM2009}. The enhancement factor is an average of the LDOS over the emission linewidth. It should be noted that a lossless resonance mode leads to a delta-function peak in the LDOS spectrum. \\
In conclusion, the PhC cavity can enhance the spontaneous emission rate of the emitter positioned in the cavity mode when the spectral, spatial and polarization overlap is fulfilled, namely matching the emission pattern of the source with the resonance mode of the cavity. In this case, the Purcell enhancement factor $F_\mathrm{Purcell}$ can be derived as the ratio of spontaneous emission rate $\Gamma$ of the emitter coupled to the resonant mode of the cavity, to the free-space emission rate $\Gamma_0$ \cite{purcellResonanceAbsorptionNuclear1946,coccioliSmallestPossibleElectromagnetic1998}
\begin{equation}
{F_\mathrm{Purcell}}=\frac{\mathrm{\Gamma}}{\mathrm{\Gamma}_0}=\frac{3}{4\pi^2}\cdot\left(\lambda^{(n)}\right)^3\cdot\frac{Q^{\left(n\right)}}{V^{\left(n\right)}}
\label{eq:purcell}
\end{equation}
where $\lambda^{(n)}=\lambda^{(n)}_{\mathrm{vac}} / n_{\mathrm{mat}}$ is the resonance wavelength of the mode containing the host material of the emitter, here diamond, with the refractive index $n_{\mathrm{mat}}$.
Equation (\ref{eq:purcell}) is valid when the linewidth of the resonance mode is broader than the transition linewidth of the emitter. This condition needs to be fulfilled when the source is positioned in the maximum of the electric field of the resonance mode and when the emitters transition is matched spectrally and in polarization \cite{coccioliSmallestPossibleElectromagnetic1998}. A tremendous progress is obtained in the development of the methods to increase the Purcell enhancement, which can be achieved either by increasing the quality factor \cite{zhangUltrahighQTETM2009,quanDeterministicDesignWavelength2011,zhouRefractiveIndexSensing2014,witmerDesignNanobeamPhotonic2016,kuramochiUltrahighQOnedimensionalPhotonic2010} or by reducing the mode volume \cite{miuraUltralowModevolumePhotonic2014}.

\subsection{Freestanding, Cross-Bar Photonic Crystal Cavities}

\subsection*{Investigation of Modulated Photonic Crystals (Bragg Mirror)}

A key challenge for many quantum optics experiments is to increase the light-matter interaction strength. On-chip PhC cavities are promising candidates to provide strong Purcell enhancement and high coupling efficiencies. PhC cavities utilize two non-uniform Bragg gratings, consisting of a number of periodic segments. In between both Bragg gratings a cavity region is introduced. In the subsequent experiments quantum emitters will be positioned in this region yielding an efficient light-matter interface. \\
The first step in engineering the PhC cavities is the determination of the Bragg mirror’s parameters, which can efficiently confine light in a small volume. The aim of computational simulations is to design the Bragg mirror with a bandgap in the required frequency region, while providing low radiation loss. The radiation losses can be minimized by modulating the refractive index of the segments. There are distinguished dielectric types of cavities, where antinodes of the major electric field component, the $E_y$-component, lies in the dielectric cavity region (in $\mathrm{Si}_3\mathrm{N}_4$). There are also air type cavity regions, where the antinodes of the major electric field component, the $E_y$-component, lie in the air (holes).  In order to obtain enhancement of the emitted light, the source should be placed in the antinode of the major electric field of the corresponding mode. Depending on the application the device needs to be designed and optimized correspondingly.

\subsubsection*{Choice of the Bragg Mirror Parameters.}
The width of the waveguide is chosen to produce a single transverse electric fundamental mode (TE0) in the wavelength range of interest. The thickness is pre-determined at $200\nm$ by the choice of the conventional silicon nitride-on-insulator wafer. The period $ a $ of the air holes can be estimated according to Bragg’s law \cite{braggReflectionXraysCrystals1913} as $a=\lambda_0/(2n_{\mathrm{eff}})$, where $n_{\mathrm{eff}}$ is the effective refractive index of the mode. When optimizing for the ZPL of the SiV center at $\lambda=738\nm$ a refractive index of $n_{\mathrm{eff}}=1.546$ needs to be considered. 
According to Bloch-Floquent theory it is sufficient to determine the bandgap for a single segment of the 1D periodic PhC, which can be allocated to the uniform PhC with infinite number of such periods. 
It should be noted, that a smooth transition of the resonance mode into the waveguide mode at the edge mirror-waveguide can be insured by modulation of the segments. Quadratic interpolation of the filling fraction $ff$ \cite{quanDeterministicDesignWavelength2011} of the Bragg mirrors segments reduces radiation losses. 
The exploration of the geometric parameters for these device are performed using the free open-source MPB software \cite{johnsonBlockiterativeFrequencydomainMethods2001}. The aim is to ensure that all the segments in the mirror cover the necessary bandgap range and the attenuation constant, the so called mirror strength \cite{quanDeterministicDesignWavelength2011}, of the inner segment is zero, while for the outer segments this values should be maximized.

\subsection*{Simulations of the Nanobeam and Cross-Bar PhC with an Internal Cavity}
\begin{figure}
\includegraphics[width=\textwidth]{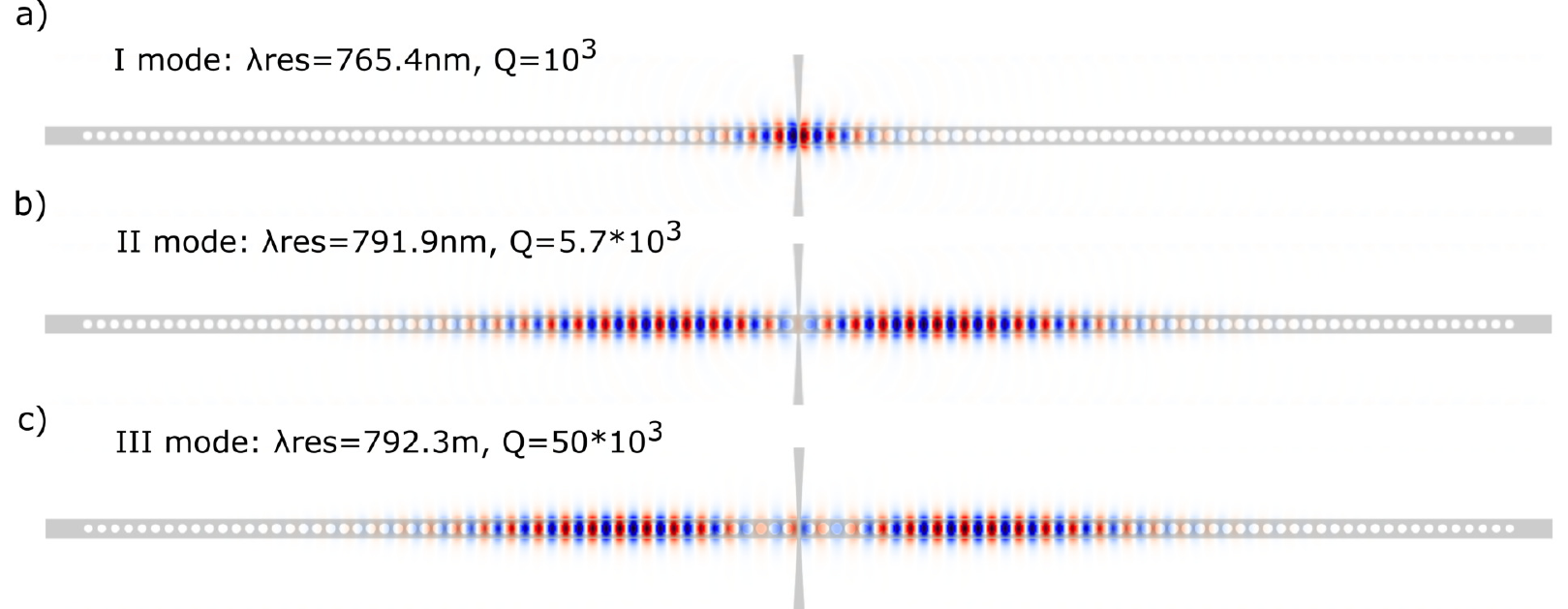}
\caption[]{Simulated Electric field distribution ($E_y$ component) for TE-like fundamental (a), second-order (b) and third-order resonance mode profiles, supported in the freestanding cross-bar PhC cavity. The electric field profiles are superimposed with the dielectric permittivity of the structure. PhC cavity parameters: cavity length $190\nm$, periodicity  $a=300\nm$ and number of holes on each side $N=53$.\\ 
	Copyright declaration: First published in the PhD Thesis of Dr. Anna P. Ovvyan \lq{}\lq{}Nanophotonic circuits for single photon emitters" DOI: 10.5445/IR/1000093929}
\label{fig:simulated_field_PCC}
\end{figure}

\begin{figure}
\includegraphics[width=\textwidth]{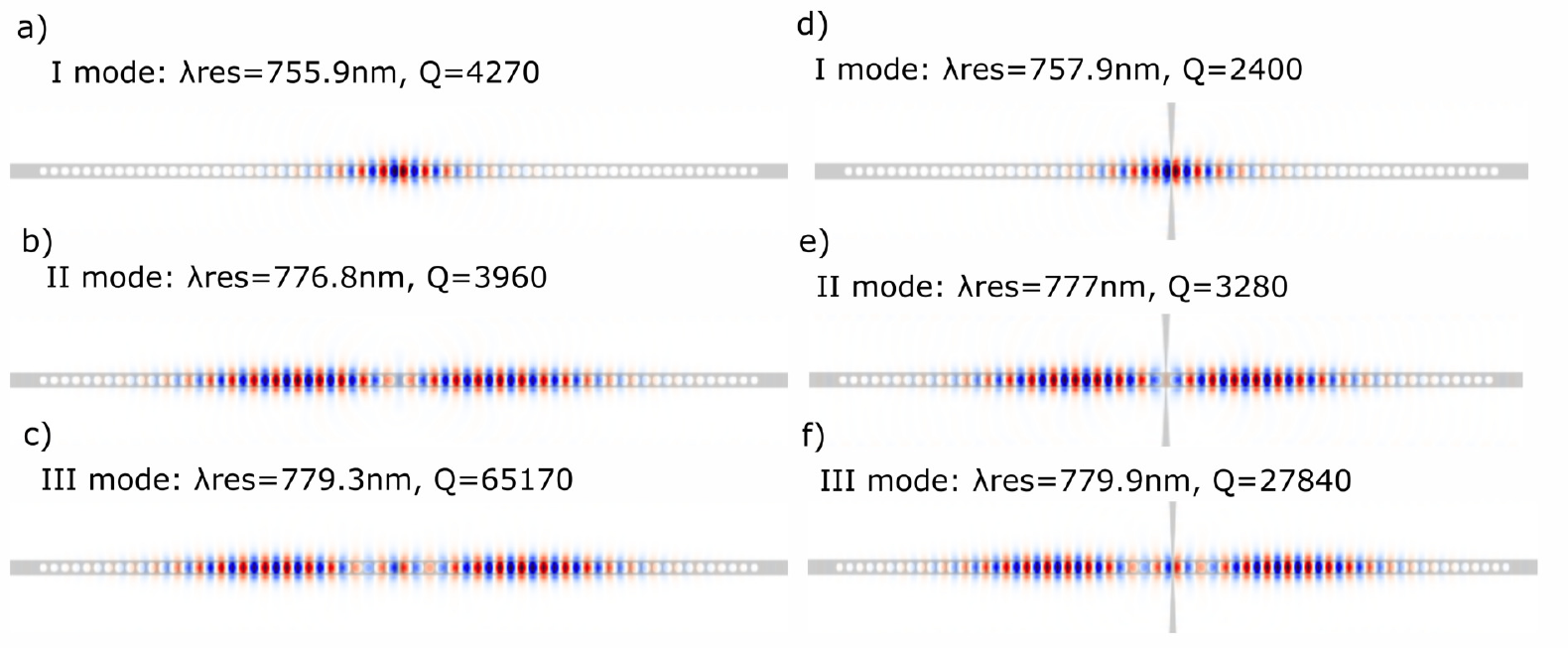}
\caption[]{Simulated Electric field distribution ($E_y$ component) for the TE-like I mode (a,d), II mode (b,e) and III mode (c,f) resonance mode profiles supported in the freestanding nanobeam PhC cavity (a,b,c) and the cross-bar PhC cavity (d,e,f). The electric field profiles are superimposed with the dielectric permittivity of the structure. PhC cavity parameters: cavity length $200\nm$, periodicity $a=290\nm$ and number of holes on each side $N=33$.\\ 
	Copyright declaration: First published in the PhD Thesis of Dr. Anna P. Ovvyan \lq{}\lq{}Nanophotonic circuits for single photon emitters" DOI: 10.5445/IR/1000093929}
\label{fig:simulated_field_nanobeam_PCC}
\end{figure}

One-dimensional PhC cavities enable the direct coupling of an optical transition to nanophotonic waveguides and thus provide compatibility to integrated optical circuits. The light-matter interaction can be strongly enhanced through the Purcell effect by placing a quantum emitter in the cavity region. 
\begin{figure}
\includegraphics[width=\textwidth]{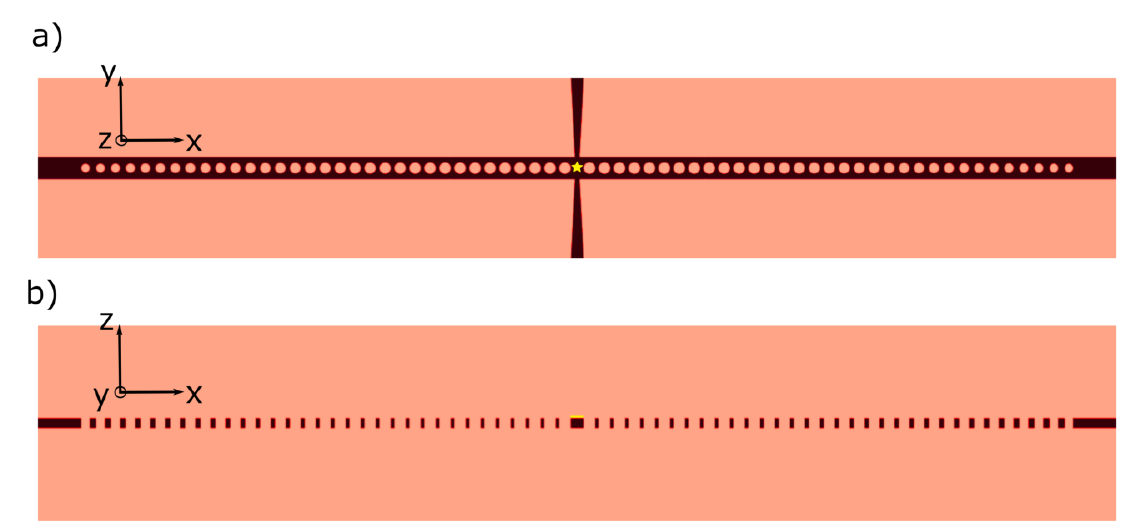}
\caption[]{Schematic of a freestanding cross-bar PhC cavity with an evanescently coupled emitter (\textit{yellow star}), placed on top of cavity region in $xy$ plane (a) and $xz$ plane shown on (b). These sketches are obtained as an output of the performed simulations using the free open-source software MEEP.\\ 
	Copyright declaration: First published in the PhD Thesis of Dr. Anna P. Ovvyan \lq{}\lq{}Nanophotonic circuits for single photon emitters" DOI: 10.5445/IR/1000093929}
\label{fig:schematic_freestanding_PCC}
\end{figure}
With the help of a novel cross-bar design (Fig. \ref{fig:schematic_freestanding_PCC}) it is possible to spatially separate the excitation (waveguide in $ y $-direction) and emission (PhC cavity in $ x $-direction). The collection efficiency and emission rate will be enhanced due to the Purcell effect. Thus, the second step of the 3D simulations is focused on the determination and optimization of high-$Q$ cavity modes with geometric parameters previously determined via the MPB simulations, as shown in the previous paragraph by employing the open-source software MEEP \cite{oskooiMeepFlexibleFreesoftware2010,tafloveAdvancesFDTDComputational2013}. The Gaussian pulse dipole source with an $E_y$ electric field is positioned in the center of the cavity with the pulse centered at $\lambda_{\mathrm{res}}$ and with a bandwidth $\Delta\lambda$. The electric field distributions of the first three resonance modes of the cross-bar PhC cavity are shown in Fig. \ref{fig:simulated_field_PCC}. The fundamental resonant mode, also referred to as the I mode, is characterized by a single central antinode envelope of the $E_y$ pattern. For the second mode, further called II mode, the envelope consists of two antinode wings. The third mode, further called III mode, shows an envelope with 3 antinode symmetric around the center. 
As can be seen from the simulations, the quality factor of the I mode is $Q=10^3$ which is 50-times smaller than for the III mode $Q=50\cdot10^3$. This can be partly attributed to the cross-bar geometry. In Fig. \ref{fig:simulated_field_nanobeam_PCC} the effect of the cross-bar design is visualized.
The $Q$-factor of odd resonance modes of the nanobeam PhC is about two times higher in comparison with the cross-bar PhC cavity, due to the presence of the crossed waveguide in the middle of the cavity, where there is an antinode of the electric field envelope of these modes. In contrast, the cross-bar has not much influence on the $Q$-factor of even modes, since the antinodes of these modes are concentrated in the wings of the cavity. Note, that the wavelengths of the corresponding resonant modes are approximately the same for both types of PhC cavities.

\subsection*{Simulation and Experimental Investigation of the PhC Cavity Parameters}

\begin{figure}
\includegraphics[width=\textwidth]{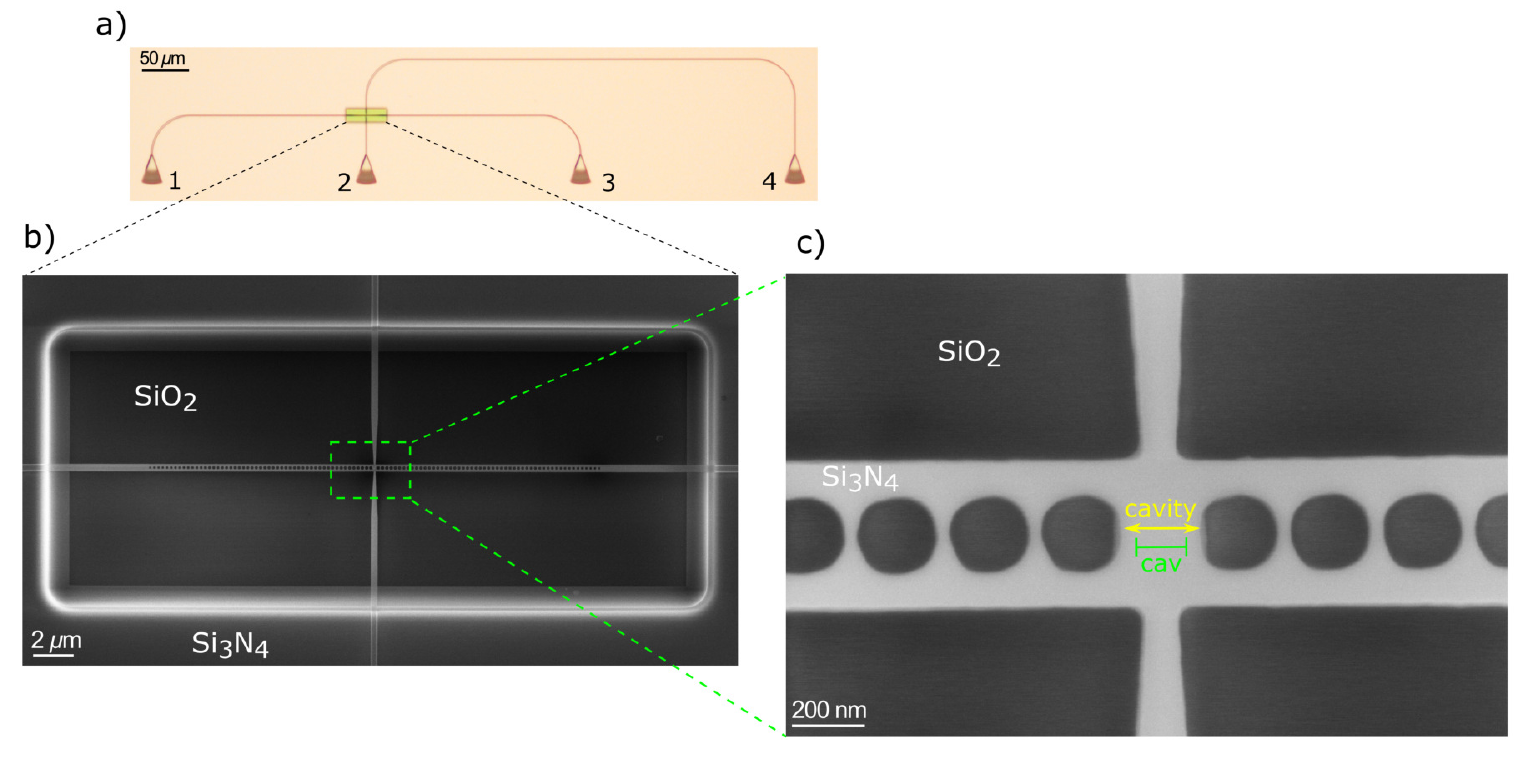}
\caption[]{Freestanding cross-bar PhC cavity. (a) Optical microscope image of a PhC cavity device. (b) SEM image of a freestanding PhC cavity with zoom-in in to the cavity region (c), where the green line (cav.) indicates the designed cavity length, i.e. the distance between two central segments.\\ 
	Copyright declaration: First published in the PhD Thesis of Dr. Anna P. Ovvyan \lq{}\lq{}Nanophotonic circuits for single photon emitters" DOI: 10.5445/IR/1000093929}
\label{fig:freestanding_PCC}
\end{figure}

\begin{figure}
\includegraphics[width=\textwidth]{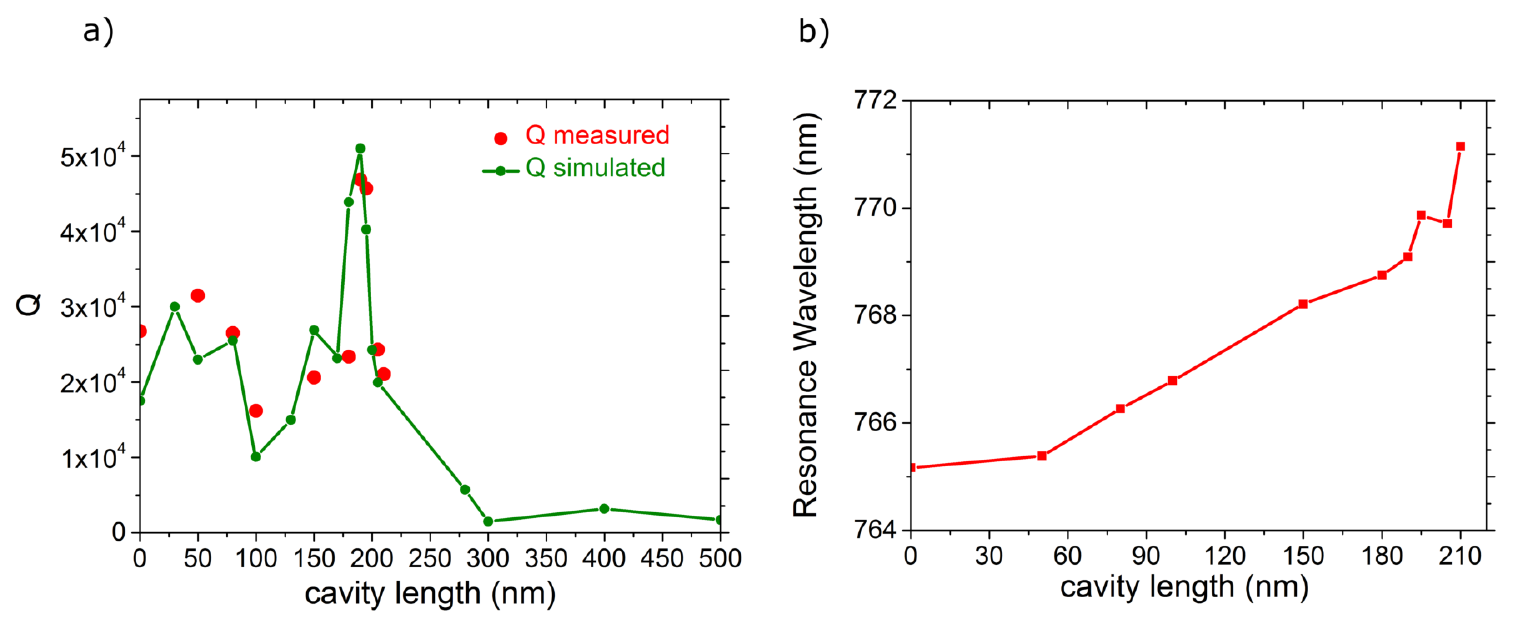}
\caption[]{Experimentally measured (\textit{red dots}) and simulated (\textit{green curve}) dependency of the (a) $Q$-factor and (b) resonance wavelength of the III-order TE-like resonance mode on the cavity length for a freestanding cross-bar PhC cavity with the following parameters: period $a=290\nm$, each modulated Bragg mirror consists $N=53$ segments with quadratic tapered holes.\\ 
	Copyright declaration: First published in the PhD Thesis of Dr. Anna P. Ovvyan \lq{}\lq{}Nanophotonic circuits for single photon emitters" DOI: 10.5445/IR/1000093929}
\label{fig:experimental_Q_freestanding_PCC}
\end{figure}

\begin{figure}
\includegraphics[width=\textwidth]{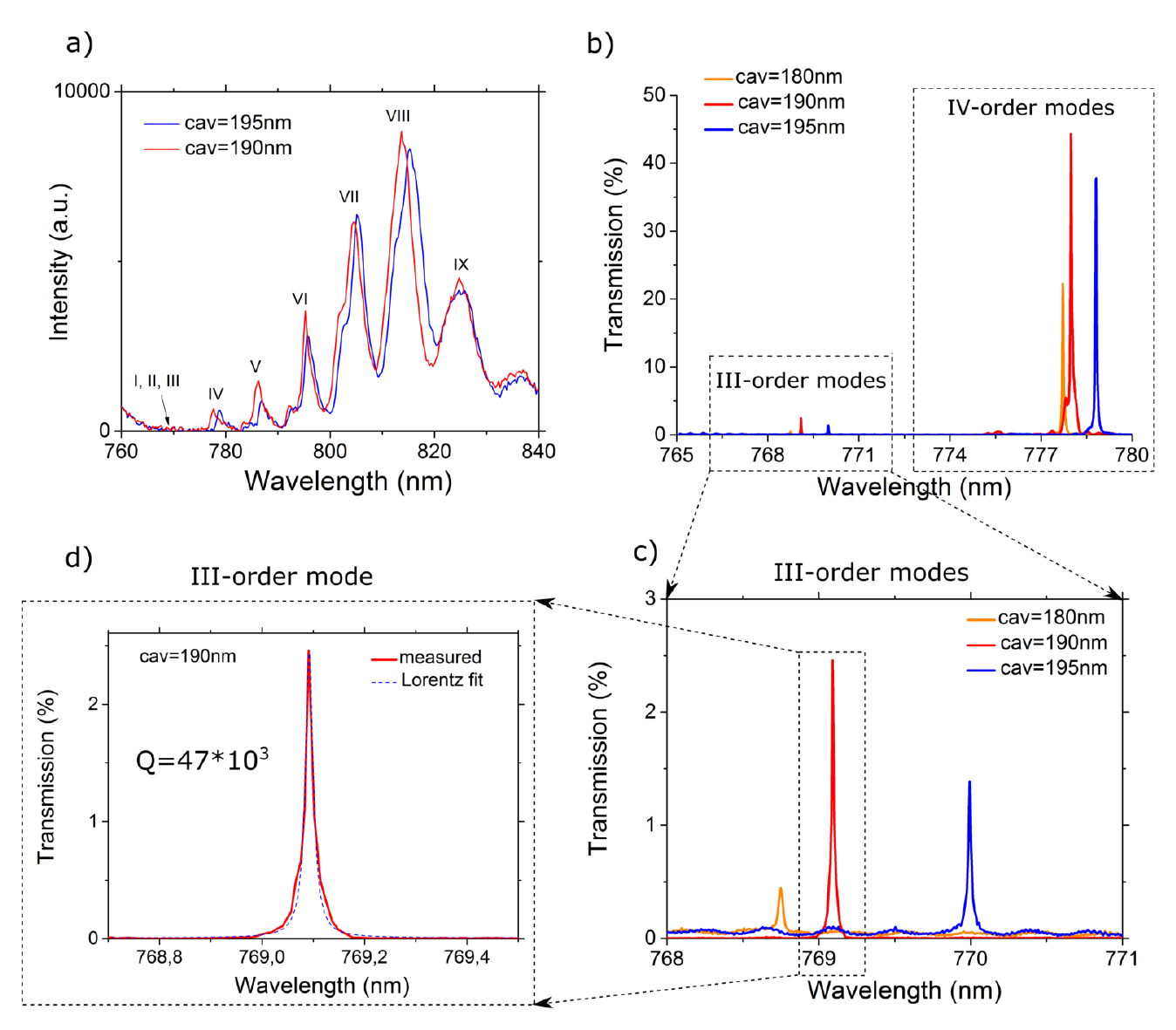}
\caption[]{Measured transmission spectrum of freestanding cross-bar PhC cavities with optimized cavity length $cav=190\nm$ (\textit{red curve}), $195\nm$ (\textit{blue curve}) by coupling supercontinuum light (a), tunable laser light in the range $765$--$780\nm$ (b) with zoom-in into III-order modes on (c). The index number of the resonance modes is annotated in the plots. (d) Measured III-order resonance peak (\textit{solid curve}) and a Lorentzian fit (\textit{dashed curve}), yielding $Q=47\cdot 10^3$. Parameters of measured PhC cavity: period $a=290\nm$, $N=53$ quadratic tapered segments.\\ 
	Copyright declaration: First published in the PhD Thesis of Dr. Anna P. Ovvyan \lq{}\lq{}Nanophotonic circuits for single photon emitters" DOI: 10.5445/IR/1000093929}
\label{fig:experimental_Q_transmission_freestanding_PCC}
\end{figure}

The fabricated freestanding, cross-bar PhC cavities are shown in Fig. \ref{fig:freestanding_PCC}. The devices are terminated with grating couplers, labeled 1 and 3, to perform transmission measurements, while the couplers labeled 2 and 4 are utilized for crosstalk measurements.
The cavity length, indicated as cav in Fig. \ref{fig:freestanding_PCC}, is the distance between two central segments and has the highest influence on the $Q$-factor. The behavior of the III-order mode is explored since it showed the highest quality factor according to the simulations. The simulation results of the $Q$-factors for the freestanding cross-bar PhC cavity in dependence on the cavity length are shown in Fig. \ref{fig:experimental_Q_freestanding_PCC}. The results are in agreement with measured $ Q $ values \cite{fehlerEfficientCouplingEnsemble2019}. The highest measured Q-factor is $Q = 47\cdot10^3$ (Figure \ref{fig:experimental_Q_freestanding_PCC} b)) at the optimal cavity length of $190\nm$, the simulated value for this device is $Q = 51\cdot10^3$. The resonance wavelength increases with an increase of the cavity length. \\
The measured transmission spectrum of the cross-bar PhC cavity with an optimized cavity length of $190\nm$ is displayed in Fig. \ref{fig:experimental_Q_transmission_freestanding_PCC}. The spectrum shown in a) is measured over a broad wavelength range using a transmission measurement with a supercontinuum light source. The low resolution ($1.3\nm$ FWHM) and low sensitivity spectrometer does not enable to detect the first three high $Q$-factor resonance modes. The employed tunable laser, with a limited tuning range of $765-780\nm$, enables to determine the III- and IV-order resonance modes in the transmitted light. The cavity resonances III- and IV-order peaks with the highest measured $Q=47\cdot 10^3$ are shown in Fig. \ref{fig:experimental_Q_transmission_freestanding_PCC}.

\subsection*{Tuning Mechanisms for Resonance Wavelength}
\begin{figure}
\includegraphics[width=0.6\textwidth]{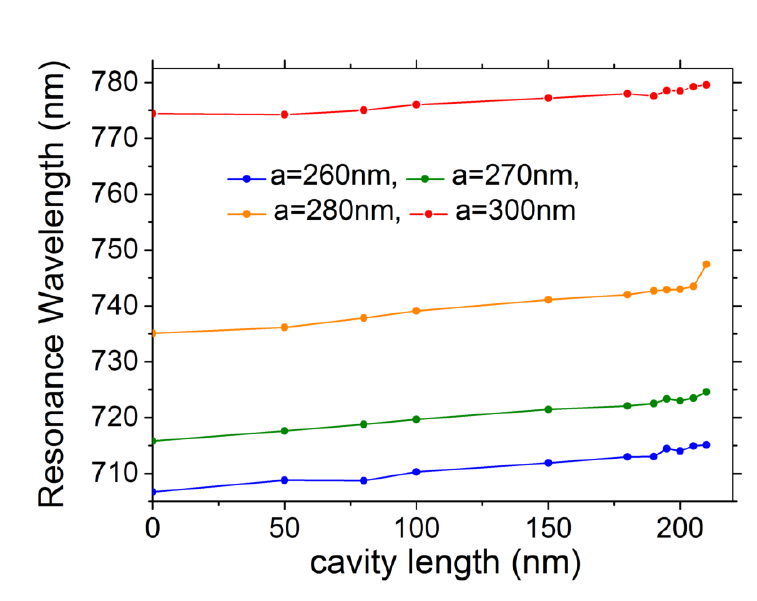}
\caption[]{Experimentally measured resonance wavelength of the I mode as function of cavity length for freestanding cross-bar PhC devices with different periods as indicated by the corresponding color and label in the legend of the plot.\\ 
	Copyright declaration: First published in the PhD Thesis of Dr. Anna P. Ovvyan \lq{}\lq{}Nanophotonic circuits for single photon emitters" DOI: 10.5445/IR/1000093929}
\label{fig:experimental_resonance_wl_freestanding_PCC}
\end{figure}

\begin{figure}
\includegraphics[width=\textwidth]{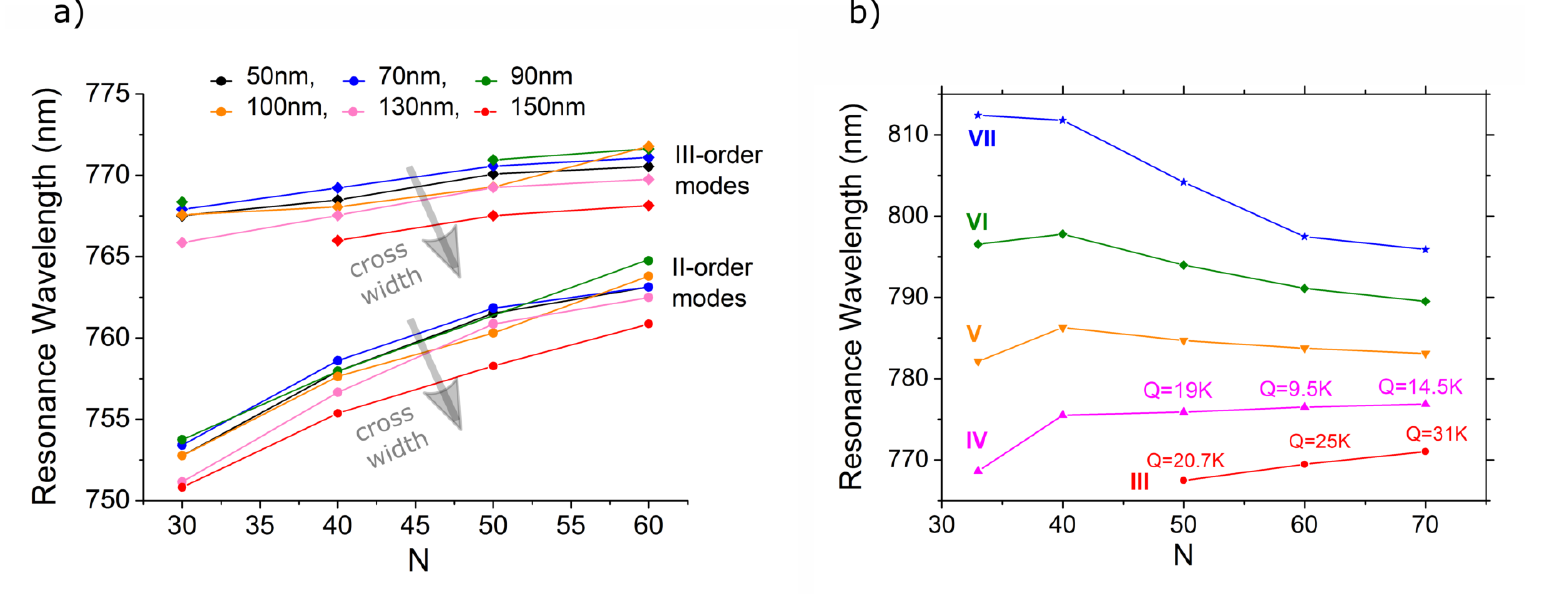}
\caption[]{(a) Experimentally measured dependency of resonance of the II- and III-order modes on the number of mirror segments N. Colorful curves correspond to freestanding PhC devices with different widths of cross-bar indicated in the legend of the plot. PhC cavity parameters: $a=295\nm$, $cav=200\nm$. (b) Resonance wavelength as a function of number of mirror segments $N$, with the mode index being indicated in the plot. PhC cavity parameters: $a=290\nm$, $cav=200\nm$, width of cross-bar $100\nm$.\\ 
	Copyright declaration: First published in the PhD Thesis of Dr. Anna P. Ovvyan \lq{}\lq{}Nanophotonic circuits for single photon emitters" DOI: 10.5445/IR/1000093929}
\label{fig:dependency_resonance_wl_segments_freestanding_PCC}
\end{figure}

Controlling the spectral position of the resonance mode is important in order to achieve the maximum coupling strength. Increasing the periodicity, $a$, leads to a rise of the resonance wavelength based on Bragg's law, which is evident in Fig. \ref{fig:experimental_resonance_wl_freestanding_PCC}. The derived tuning coefficient of the resonance wavelength with the periodicity amounts as $\frac{\delta\lambda}{\delta\Lambda}\approx0.5$--$2$.
Furthermore, an increasing cavity length leads to an increase of the resonance wavelength. The derived tuning coefficient of the resonance wavelength with the cavity length amounts as $\frac{\delta\lambda}{\delta{}cav}\approx0.02$--$0.05$.\\
Changing the periodicity and cavity length can be utilized as tuning mechanisms for the resonance wavelength of the developed cross-bar PhC cavity, unless a too drastic change of these parameters shifts the bandgap out of the spectral range of interest.\\
It should be noted, that the width of the tapered end of the cross-bar and the number of mirror segments of the PhC cavity also impacts the resonance wavelength of the mode as experimentally investigated in Fig. \ref{fig:dependency_resonance_wl_segments_freestanding_PCC}. The increase of the width of the crossed waveguides leads to an increase of the effective refractive index in the middle of the cavity and a shift the resonance towards longer wavelength would be expected. Surprisingly, it leads to decrease of the resonance wavelength. This can be explained in the following way: An increase of the tapering width of the cross-bar leads to a rise of the leakage of the resonance mode into the crossed waveguide for each cycle of the propagation of the mode inside the cavity, which leads to an increase of the crosstalk. Therefore, the penetration of the mode inside the PhC Bragg mirrors is slightly lower, resulting in a reduction of the effective length of the cavity, assuming that other parameters are kept constant. The resulting decrease of the resonance wavelength is illustrated in Fig. \ref{fig:dependency_resonance_wl_segments_freestanding_PCC} a). Indeed, increasing the width of the crossed waveguide leads to a decrease of the $Q$-factor due to the higher value of the crosstalk.\\
Another experimentally investigated trend is illustrated in Fig. \ref{fig:dependency_resonance_wl_segments_freestanding_PCC} b). An increase of the number of mirror segments elevates the resonance wavelength for II- and III-order modes because of the increase of the effective length of PhC cavity, as a result of a deeper penetration of the mode into the Bragg mirrors from both sides of the cavity. At the same time, for higher order modes this trend is not always maintained, as indicated in Fig. \ref{fig:dependency_resonance_wl_segments_freestanding_PCC} b). Namely, higher order modes with lower $Q$-factor experience higher radiation loss. Starting from a certain $N$, a further increase of the number of segments yields a shorter penetration depth of the resonance mode into the Bragg mirror, leading to a reduction of the effective length of the PhC cavity and a decrease in resonance wavelength. As shown in Fig. \ref{fig:dependency_resonance_wl_segments_freestanding_PCC} b) the effect of shifting the resonance wavelength is stronger for higher index modes. For III- and IV-order modes, increasing $N$ still leads to a slight rise of the resonance wavelength. In contrast, with V-, VI- and VII- modes the resonance shift towards lower wavelength range starts from $N=40$ mirror segments.

\section{Color Center in Nanodiamonds}

\begin{figure}
\includegraphics[width=\textwidth]{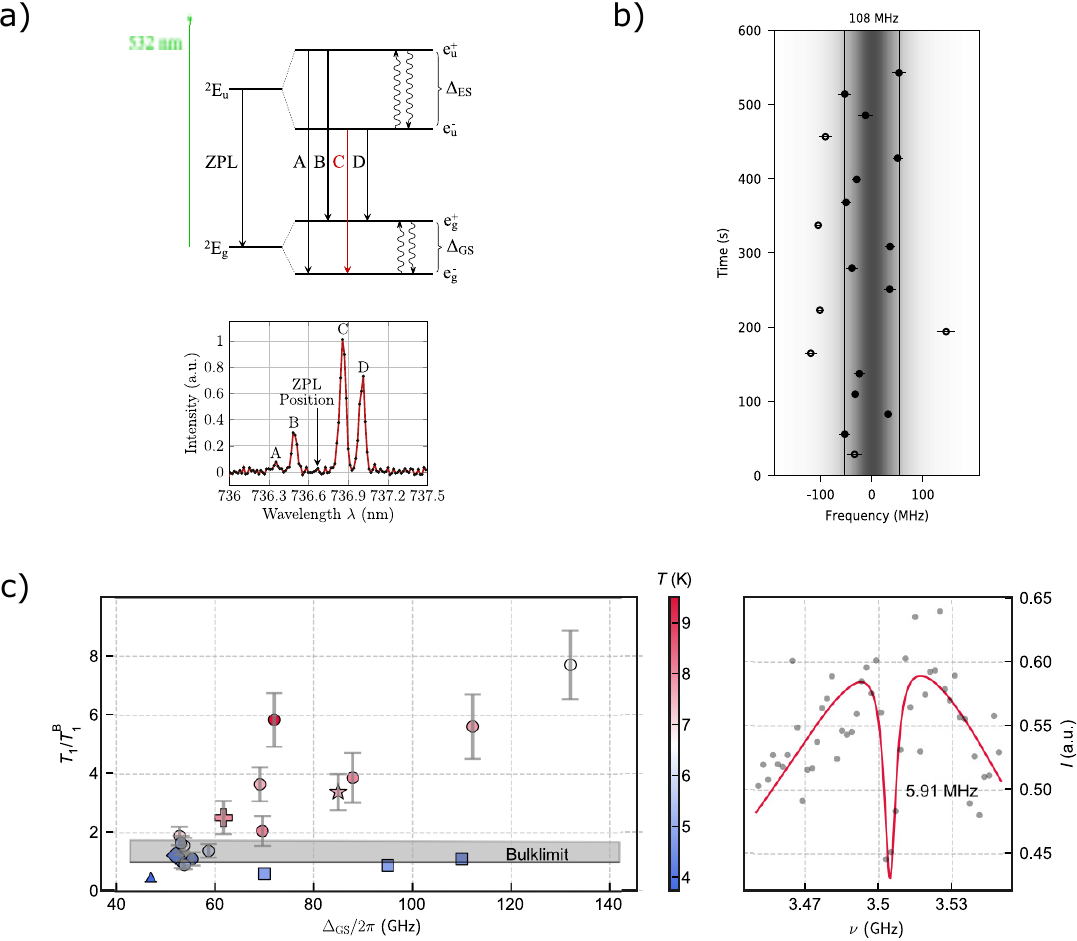}
\caption[]{\textbf{Color center in nanodiamond.} (a) The energy level fine structure of the SiV center in diamond (\textit{upper panel}). The finestructure was recently resolved in NDs (\textit{lower panel}). (b) The spectral stability is measured in photoluminescence excitation (PLE) spectroscopy over 10 minutes and reveals that the central PLE peak position remains within the Fourier-Transform limit, here 108 MHz. (c) The spin properties of SiV center in NDs can be prolonged as compared to SiV center in bulk diamond due to a change in the phononic density of states. The orbital relaxation rate (\textit{left panel}), disclosed by the orbital $T_1$-time, is measured to be up to 8-times longer than the best possible value for bulk diamond at similar circumstances. Respectively, the spin dephasing time $T_2^*$ is also prolonged in NDs (\textit{right panel}). $T_2^*$ is inferred from CPT measurements where the linewidth of the CPT dip extrapolated to zero power yields $T_2^*$.\\
Copyright declaration: (a) Reprinted figure with permission from Häußler, S., et al. New J. Phys 21, 103047, 2019. Copyright (2019) author(s), licensed under https://creativecommons.org/licenses/by/3.0/legalcode. (b) Reprinted figure with permission from Rogers, L. J., et al. Phys. Rev. Applied, 11, 024073, 2019. Copyright (2019) by the American Physical Society. (c) Reprinted figure with permission from Klotz, M., et al. Phys. Rev. Lett., 128, 153602, 2022. Copyright (2022) by the American Physical Society.}
\label{fig:ND}
\end{figure}

In this section, we describe color centers in diamond hosts with a size small enough to enable hybrid integration. In case of optical interfacing, the size needs to be small as compared to the wavelength of the coupled optical transition. The diamond host could be tailored by nanofabrication towards specific applications. For example, nanocone structures for directional, enhanced light emission have been fabricated and are also capable for the pick-and-place method, here done via tungsten tip \cite{jeonBrightNitrogenVacancyCenters2020}. Therefore, an integration into hybrid quantum photonics devices is possible. However, in the following we focus on NDs that do not require any a priori diamond nanostructuring. NDs have a long research history and a broad application range \cite{alkahtaniFluorescentNanodiamondsPresent,qinNanodiamondsSynthesisProperties2021,bradacRoomtemperatureSpontaneousSuperradiance2017,juanCooperativelyEnhancedDipole2017}. Early studies focused on NV centers in NDs demonstrating, for example, single spin manipulation via ODMR \cite{tislerFluorescenceSpinProperties2009}. However, the spectral instabilities and uncontrolled fluorescence blinking originating from the close proximity to the diamond surface as well as from low crystalline diamond quality remained a long-standing challenge. Also, the deterministic creation of single color centers per one ND is an ongoing challenge. It was the spectral robustness of symmetry-protected group-IV color center that enabled to overcome above mentioned limitations. Even smallest nanodiamonds as small as 1.6 nm, corresponding to approximately 400 carbon atoms, were shown to be capable to host SiV center with stable but still blinking fluorescence \cite{vlasovMolecularsizedFluorescentNanodiamonds2014} and single SiV center were demonstrated in NDs of about 10 nm in size \cite{bolshedvorskiiSingleSiliconVacancy2019}. Narrowband, bright, single photon emission from SiV centers in NDs was shown in NDs with high crystalline quality. In early work, the NDs were produced by bead assisted sonic disintegration of polycrystalline chemical vapor deposition films \cite{neuNarrowbandFluorescentNanodiamonds2011}. Later, the color center were generated by ion implantation into preselected NDs and high-temperature annealing was employed to reduce the spectral linewidth \cite{takashimaCreationSiliconVacancy2021}. Spectrally stable single photon emission even in smallest NDs with a spectral width given by the excited state lifetime, the so-called Fourier-Transform limit, became possible \cite{jantzenNanodiamondsCarryingSiliconvacancy2016}. The ND-production \cite{neuSinglePhotonEmission2011} and treatment was further improved over the years. The four-line, finestructure of single SiV center in NDs was observed for the first time in chemical vapour deposition NDs on iridium\cite{neuSinglePhotonEmission2011} and later in low-strained, surface-terminated NDs \cite{rogersSingleSiVCenters2019}, as shown in Fig. \ref{fig:ND} a). Spectral diffusion on individual lines was basically absent and the linewidth remained close to the Fourier-limit as well as the peak position remained at a fixed frequency, as depicted in Fig. \ref{fig:ND} b).\\
But not only the optical properties in NDs improved over the past years. Also the spin properties can be enhanced to a level beyond what is possible in bulk diamond. The spin dephasing time for SiV in bulk diamond is limited by rapid orbital relaxation which is caused by a resonant phonon process\cite{jahnkeElectronPhononProcesses2015}. In NDs, the orbital relaxation can be prolonged by locally modifying the phonon density of states in the relevant frequency range. For NDs of size smaller than about $100\nm$ an extended orbital relaxation rate was recently demonstrated. The orbital $T_1$-time is prolonged by up to one order of magnitude as compared to the best achievable $T_1$-time in bulk diamond \cite{klotzProlongedOrbitalRelaxation2022}, see Fig. \ref{fig:ND} c) left panel. Correspondingly, the spin dephasing time was also prolonged, as inferred from the dip width in coherent population trapping (CPT) experiments \cite{klotzProlongedOrbitalRelaxation2022}, see Fig. \ref{fig:ND} c) right panel. In conclusion, group-IV color center in NDs and, in particular, the SiV center in NDs became an attractive quantum system with improved optical and spin properties at moderate temperatures of a few Kelvin.\\ 
At the same time, the NDs can be manipulated and positioned by means of a atomic force microscope (AFM) on the nanometer-scale without loss of their optical properties. These are the key properties to enable integration into hybrid quantum photonics. With respect to external coupling, for example by means of optical integration, a unique feature is the capability to optimize all degrees of freedom of the optical coupling term including positioning and dipole orientation\cite{hausslerPreparingSingleSiV2019a}. In the next chapter, the detailed aspects of QPP are discussed.

\section{Quantum-Postprocessing} 

\begin{figure}
\includegraphics[width=\textwidth]{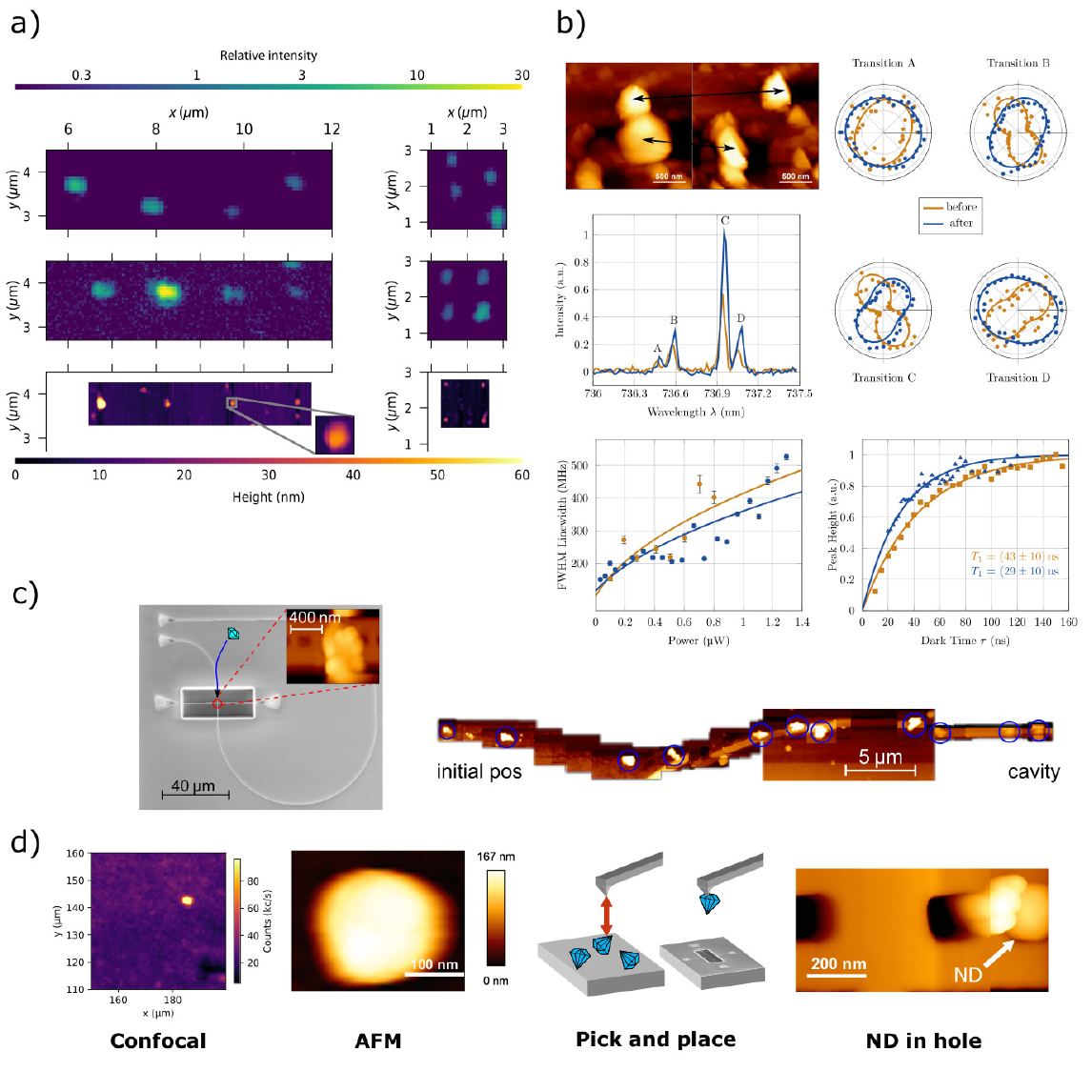}
\caption[]{\textbf{The quantum postprocessing.} (a) Confocal (\textit{upper two panels}) and AFM (\textit{lower panel}) images of randomly distributed, color centers containing NDs which are put into regular order of a 1D string (\textit{left}) and a 2D array (\textit{right}). (b) During QPP, properties such as the fine-structure, the optical linewidth and the spin $T_1$-times remain unchanged as demonstrated by the example of a rotation of an SiV containing ND. The polarization patterns change according to the rotation. (c) Demonstration of the device postprocessing via the pushing technique. A precharacterized ND is pushed via the AFM nanomanipulator, operated in contact mode, along several tens of $\umu$m in order to reach the interaction zone of the PhC cavity. (d) Demonstration of the device postprocessing via the pick and place technique. Color centers are precharacterized in cryogenic, confocal microscopy (\textit{left panel}) and AFM microscopy (\textit{second left panel}). Well-suited NDs are then picked and placed in the interaction zone of the PhC cavity. Here, the ND is placed with high precision inside the hole of the PhC cavity.\\
Copyright declaration: (a) Reprinted figure with permission from Rogers, L. J., et al. Phys. Rev. Applied, 11, 024073, 2019. Copyright (2019) by the American Physical Society. (b) Reprinted figure with permission from Häußler, S., et al. New J. Phys 21, 103047, 2019. Copyright (2019) author(s), licensed under https://creativecommons.org/licenses/by/3.0/legalcode. (c) Adapted figure with permission from Fehler et al., Nanophotonics, 9(11), 3655-3662, 2020. Copyright (2020) author(s), licensed under https://creativecommons.org/licenses/by/4.0/legalcode. (d) Reprinted with permission from Fehler et al., ACS Photonics 8(9), 2635–2641, 2021. Copyright (2021) author(s), licensed under https://creativecommons.org/licenses/by-nc-nd/4.0/legalcode}
\label{fig:QPP}
\end{figure}

The emitter-to-structure approach builds on QPP that needs to be precise and enable high-throughput processing  \cite{schrinnerIntegrationDiamondBasedQuantum2020}. In pioneering work, single NDs containing NV center were placed with about 100-nanometer-scale accuracy in the near-field of a $\mathrm{SiO}_2$-microdisk resonator \cite{barclayCoherentInterferenceEffects2009}. The nanomanipulation was done by using a fiber taper. In other early work, NDs were placed on the end facet on an optical fiber positioned within the core region of the fiber \cite{ampem-lassenNanomanipulationDiamondbasedSingle2009} in order to detect the NV fluorescence directly through the fiber\cite{schroderFiberIntegratedDiamondBasedSingle2011}. Nowadays, the most precise method is established by using AFM-based nanomanipulation capable to realize assemblies with nanometer precision. Pioneering work was done using home-build nanomanipulators arranging single-NV-containing NDs \cite{vandersarNanopositioningDiamondNanocrystal2009}. On the one hand, methods which rely on positioning individual NDs have a lower throughput as compared to, for example, methods that build on covalent bounding \cite{kianiniaRobustDirectedAssembly2016}. On the other hand, these methods have a very high precision and enable, in principle, a yield of $100\,\%$. Fig. \ref{fig:QPP} a) depicts the assembly of linear chains (left) or 2D-arrangements (right) of color center containing NDs by means of AFM-nanomanipulation. The positioning is monitored with high-precision by mapping the topology of the corresponding ND with an AFM (upper panel) and by mapping the emitter position by fluorescence detection (lower panel). As a prerequisite, the properties of the color center in the NDs must persist the QPP \cite{hausslerPreparingSingleSiV2019a}. As an example, the rotation of a ND can be monitored with high accuracy by measuring the change in polarization of the four fine-structure transitions of a SiV center located inside the ND, as depicted in Fig. \ref{fig:QPP} b). Other important properties, such as the optical linewidth, the fine-structure splitting and the orbital relaxation rate remain unaffected from the nanomanipulation procedure. Therefore, pre-characterized NDs can be pushed into pre-defined interaction regions and furthermore optimized for ideal coupling, by means of lateral and dipole alignment. In pioneering work, a tapered silica fiber is used to create a small reservoir of NDs close to the target position, the so-called dip-pen technique \cite{andersonNearfieldSurfaceEnhanced2008}. Then, in the final QPP-step, the NDs are moved into the interaction zone by pushing the cantilever tip in contact mode against the ND. During the whole QPP-step the ND is not lifted from the target substrate. With this method, highly accurate QPP of a planar PhC double-heterostructure cavity was realized \cite{barthControlledCouplingSinglediamond2009}. Similarly, color center containing NDs were pushed over distances of tens of $\umu$m into the interaction region of a free-standing PhC, as depicted in  Fig. \ref{fig:QPP} c) \cite{fehlerEfficientCouplingEnsemble2019,fehlerPurcellenhancedEmissionIndividual2020}.\\
The picking and placing of nanoparticles via AFM was established a decade ago. The method was used to place NV centers in a ND inside a gallium phosphide PhC membrane \cite{schellScanningProbebasedPickandplace2011}. The pick and place technique was then further developed enabling high-precision positioning and rotation of NDs. The high spatial precision of the AFM-based nanomanipulation enables an extremely high accuracy in assembling hybrid devices. As a recent example, individual NDs hosting SiV centers were positioned inside the hole of a PhC cavity \cite{fehlerHybridQuantumPhotonics2021}, as summarized in Fig. \ref{fig:QPP} d). The technological advancements arising from the high-precision positioning are discussed in the next section.
A remaining bottleneck is the slow throughput of the QPP step. The whole QPP procedure requires a proper pre-characterization, a highly-accurate positioning step followed by a post-characterization. A large boost in efficiency can be expected from incorporating machine learning ideas. Recently, a 100-fold speedup in single photon emitter classification via second-order autocorrelation was demonstrated \cite{kudyshevRapidClassificationQuantum2020}.

\section{Hybrid Quantum Devices Based on $\mathrm{Si}_3\mathrm{N}_4$-Photonics Post-Processed with Color Center in Nanodiamonds}

In this last section, we summarize recent developments in hybrid quantum photonics building on $\mathrm{Si}_3\mathrm{N}_4$-photonics devices, as outlined in section 4, and postprocessed with color center in NDs, as outlined in section 5 and 6. In particular, cross-bar PhC cavities are discussed. We discriminate two different strategies to establish hybrid quantum photonics devices. First, optical interfacing achieved by evanescent coupling. This method was used extensively for different systems in the past. Yet, it is inherently limited in the achievable coupling strength. Second, the optical interfacing by embedded coupling. This method puts heavy demands on the QPP-step but offers ideal performance. Before we discuss both strategies we give a full characterization of the bare cross-bar PhC cavities.

\subsection{ Full Characterization of a Cross-Bar PhC Cavity for an Evanescently Coupled Emitter}

In this subsection, we discuss the analysis of most important parameters of the cross-bar PhC cavities which were developed and used for the experimental realization of hybrid quantum photonic circuits with integrated color centers in NDs.
 
\subsection*{Analysis of the Fundamental Mode of the Cross-Bar PhC Cavity}

\begin{figure}
\includegraphics[width=\textwidth]{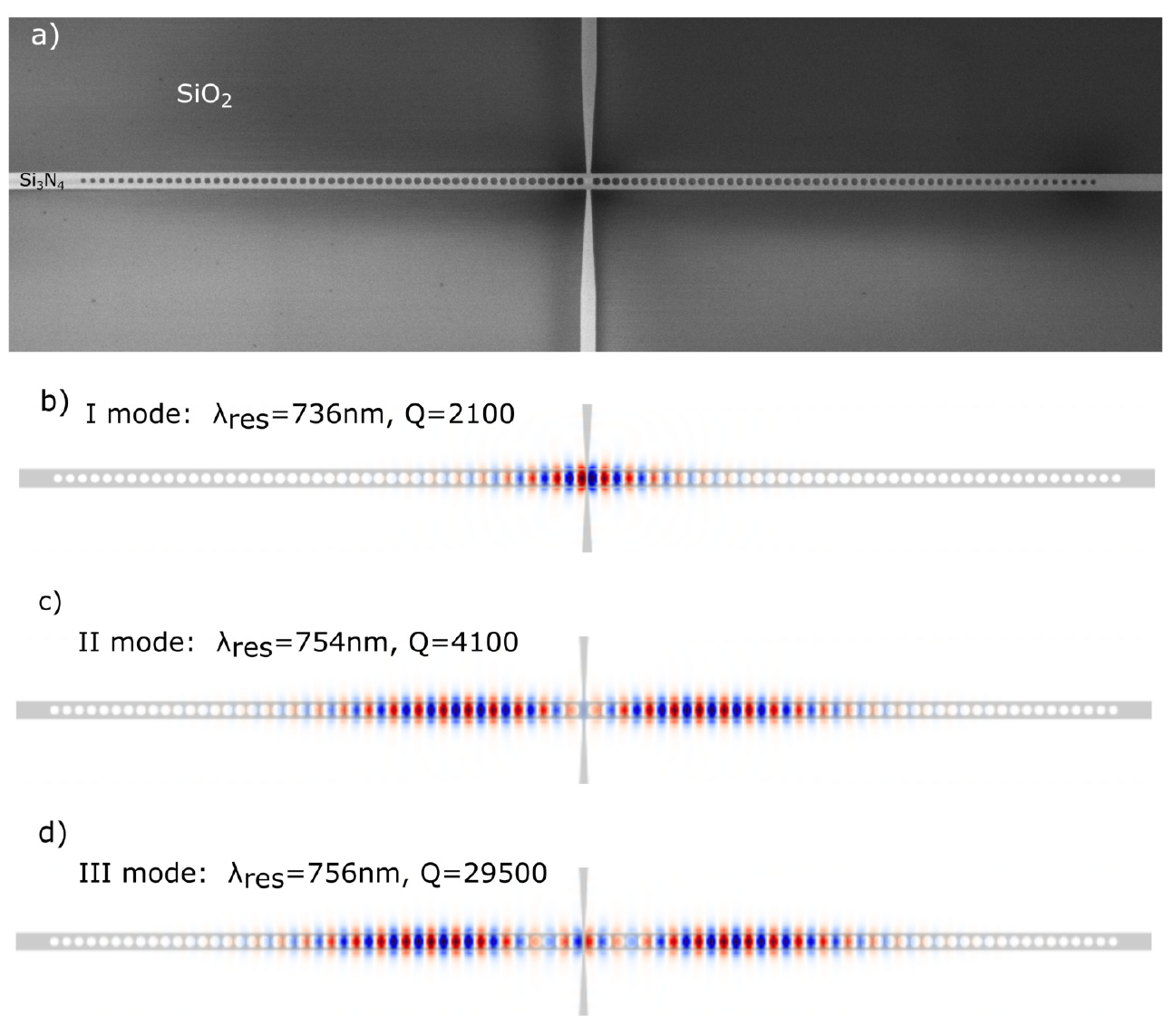}
\caption[]{(a) SEM image of a freestanding cross-bar PhC cavity. Simulated Electric field profile ($E_y$ component) of the TE-like resonance modes, superimposed with the cross-bar PhC cavity's dielectric permittivity for the I-, II-, III-order resonance modes in part (b--d) respectively.
The resonance wavelength and $Q$-factor are annotated on the plot. PhC cavity parameters: cavity length $195\nm$, periodicity $a=280\nm$, quadratic tapered $N=43$ segments.\\ 
Copyright declaration: First published in the PhD Thesis of Dr. Anna P. Ovvyan \lq{}\lq{}Nanophotonic circuits for single photon emitters" DOI: 10.5445/IR/1000093929}
\label{fig:SEM_freestanding}
\end{figure}

\begin{figure}
\includegraphics[width=\textwidth]{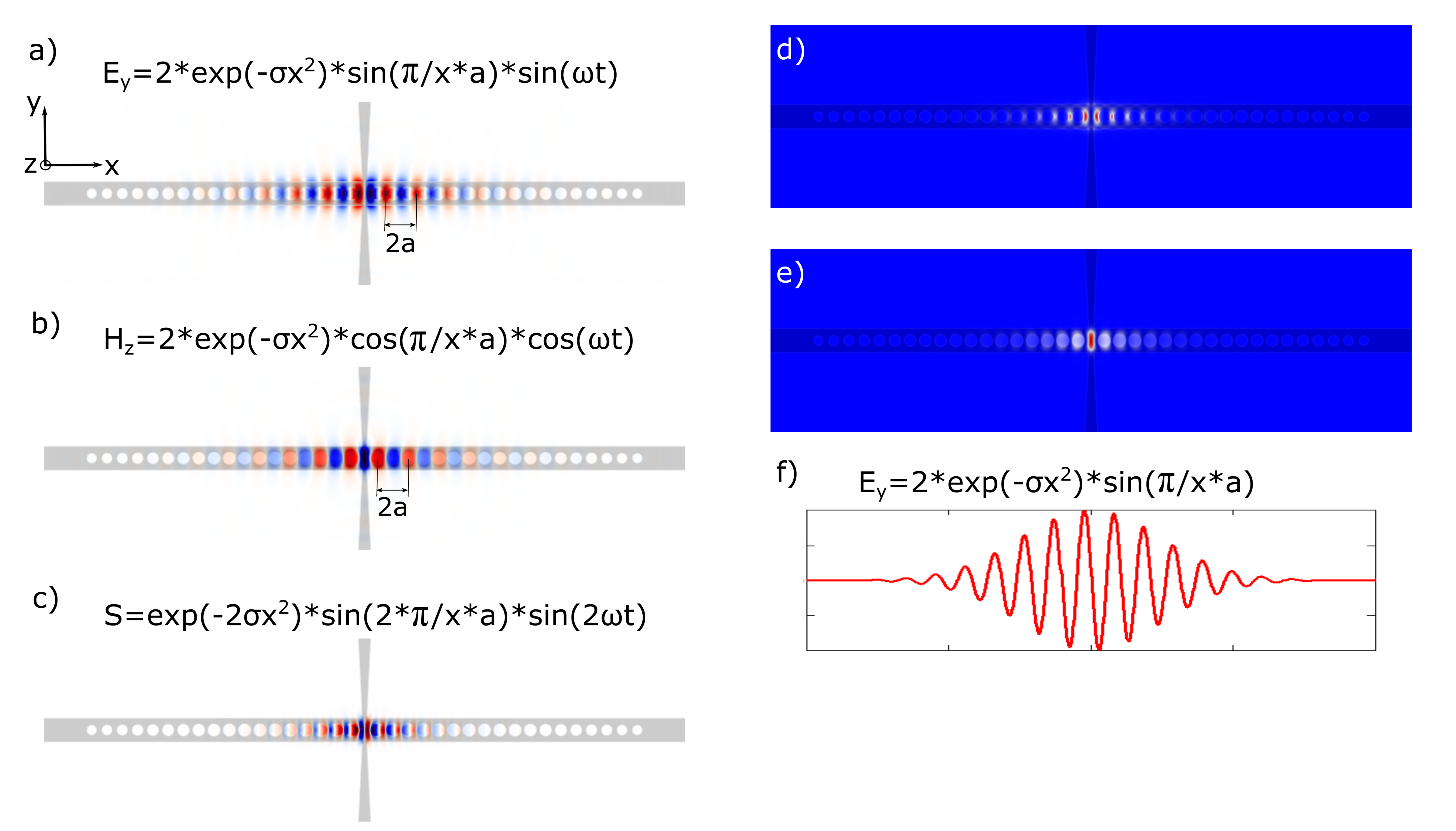}
\caption[]{Simulation results. (a) Electric field distribution ($ E_y $ component). (b) Magnetic field distribution ($ H_z $ component). (c) Poynting vector distribution. (d) Energy density distribution at the moment of time when electric energy density component is maximum. (e) Energy density distribution at the moment of time when magnetic energy density component is maximum. (f) Analytical Electric field distribution ($ E_y $ component). All distributions, superimposed with dielectric permittivity of the structure, of the fundamental TE-like resonance mode for freestanding cross-bar PhC cavity with the following parameters: cavity length $ 195\nm $, periodicity  $ a = 280\nm $, quadartic tapered $ N=43 $ segments.\\ 
Copyright declaration: First published in the PhD Thesis of Dr. Anna P. Ovvyan \lq{}\lq{}Nanophotonic circuits for single photon emitters" DOI: 10.5445/IR/1000093929}
\label{fig:simulation_freestanding_PCC}
\end{figure}

The cross-bar PhC cavities were developed to enhance the light-matter interaction, in particular for the SiV emission with a ZPL at $\lambda=738\nm$, and optimized for evanescent coupling to the fundamental modes as shown in Fig. \ref{fig:SEM_freestanding}. Further fine tuning of the resonance wavelength can be obtained by slight variation of the Bragg mirror segments' period, as discussed in section 4.
The first mode is further explored and its magnetic field ($H_z$-component), Poynting vector $S$ and energy density distributions are determined via extra cycles of three dimensional Finite Difference Time Domain (3D-FDTD) simulations. The results are displayed in Fig. \ref{fig:simulation_freestanding_PCC}. However, these simulations are carried out with a reduced number of mirror segments to decrease computation time.
Important aspects of the explored electro-magnetic field, Poynting vector and energy distributions are highlighted for the fundamental resonance mode of the cross-bar PhC cavity which are evident in Fig. \ref{fig:simulation_freestanding_PCC}.

\begin{itemize}
\item The Gaussian envelope $E_y=\sin\left(\frac{\pi x}{a}\right)\cdot\sin{\left(\omega t\right)}\cdot{\exp{\left(-\sigma x^2\right)}}$ of the electric field pattern of the I-order mode is the result of quadratic tapering of the filling fraction of the segments in each Bragg mirror from both sides of the cavity. This leads to a suppression of the mismatch between the cavity and waveguide modes and is ensured by a smooth transition of the effective refractive index of the latter. This decreases the radiation loss due to the suppression of the tails of the Fourier harmonics of the resonance mode inside the light cone.
\item The electric $E_y$ and magnetic $H_z$ field components are reversed, i.e., there is an antinode in the $H_z$ distribution in the center of cavity as opposed to the $E_y$ pattern. This can be explained by the nature of the standing wave, where the energy is passed back and forth between the region where the electric field has a maximum (magnetic field has a node) and the region where the magnetic field has a maximum (electric field has a node) with zero energy net stream in both directions, as shown in Fig. \ref{fig:simulation_freestanding_PCC} d), e). The standing wave does not transfer energy, since it is the superposition of two waves traveling in opposite directions, i.e. carry energy in opposite directions.
\item The Poynting vector of the standing wave 
\begin{equation}
S_x=E_y\cdot H_z=\sin{\left(2\omega t\right)\cdot\sin{\left(2kx\right)\cdot\exp{\left(-2\sigma x^2\right)}}}
\end{equation}
oscillates at twice the frequency of the corresponding $E_y$ and $H_z$ field components. It is important that the Poynting vector shows an impact of both magnetic and electric components at one time, in contrast with the energy distribution for the standing wave; the time-average of the Poynting vector is zero.
\item The envelope of the energy and the Poynting vector distributions described by $\exp{\left(-2\sigma x^2\right)}$ decays two times faster in comparison with the electric and magnetic fields.
\end{itemize}

\subsection*{The Spatial Map of the Local Density of States Enhancement}
\begin{figure}
\includegraphics[width=\textwidth]{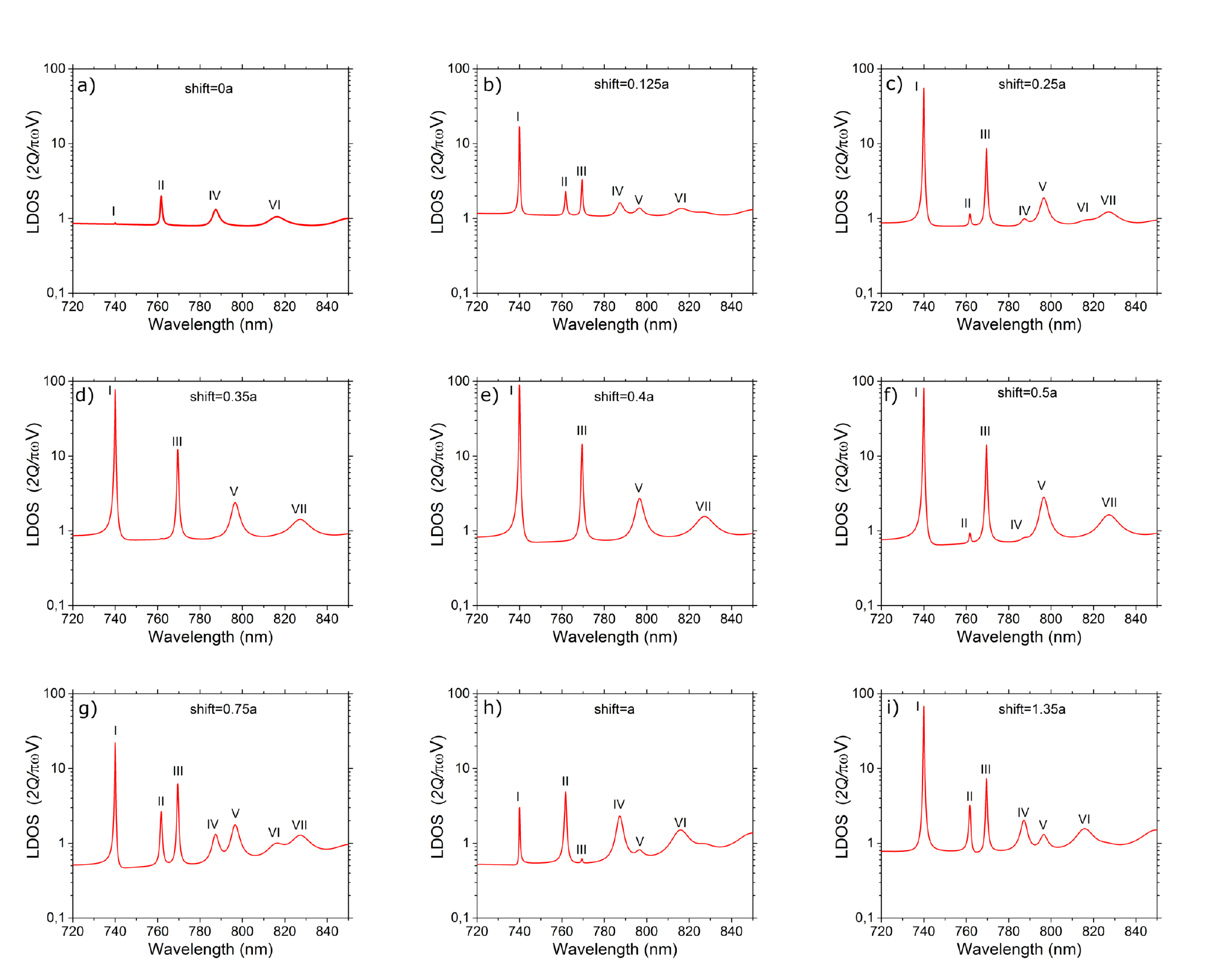}
\caption[]{Simulated LDOS spectrum enhancement of the light emitted from a dipole placed on top of the freestanding cross-bar PhC cavity, where the dipole position along the longitudinal direction is varied (a--i). The positions are annotated in terms of dipole shift from the center along the $x$-axis in Fig. \ref{fig:schematic_freestanding_PCC}, weighted in units of $a$. The freestanding cross-bar PhC cavity has the following parameters: cavity length $ 195\nm $, quadartic tapered $ N=18$ segments, periodicity  $ a = 280\nm $,. Analytical Electric field distribution ($ E_y $ component).\\ 
	Copyright declaration: First published in the PhD Thesis of Dr. Anna P. Ovvyan \lq{}\lq{}Nanophotonic circuits for single photon emitters" DOI: 10.5445/IR/1000093929}
\label{fig:simulated_ldos_spectrum}
\end{figure}

\begin{figure}
\includegraphics[width=\textwidth]{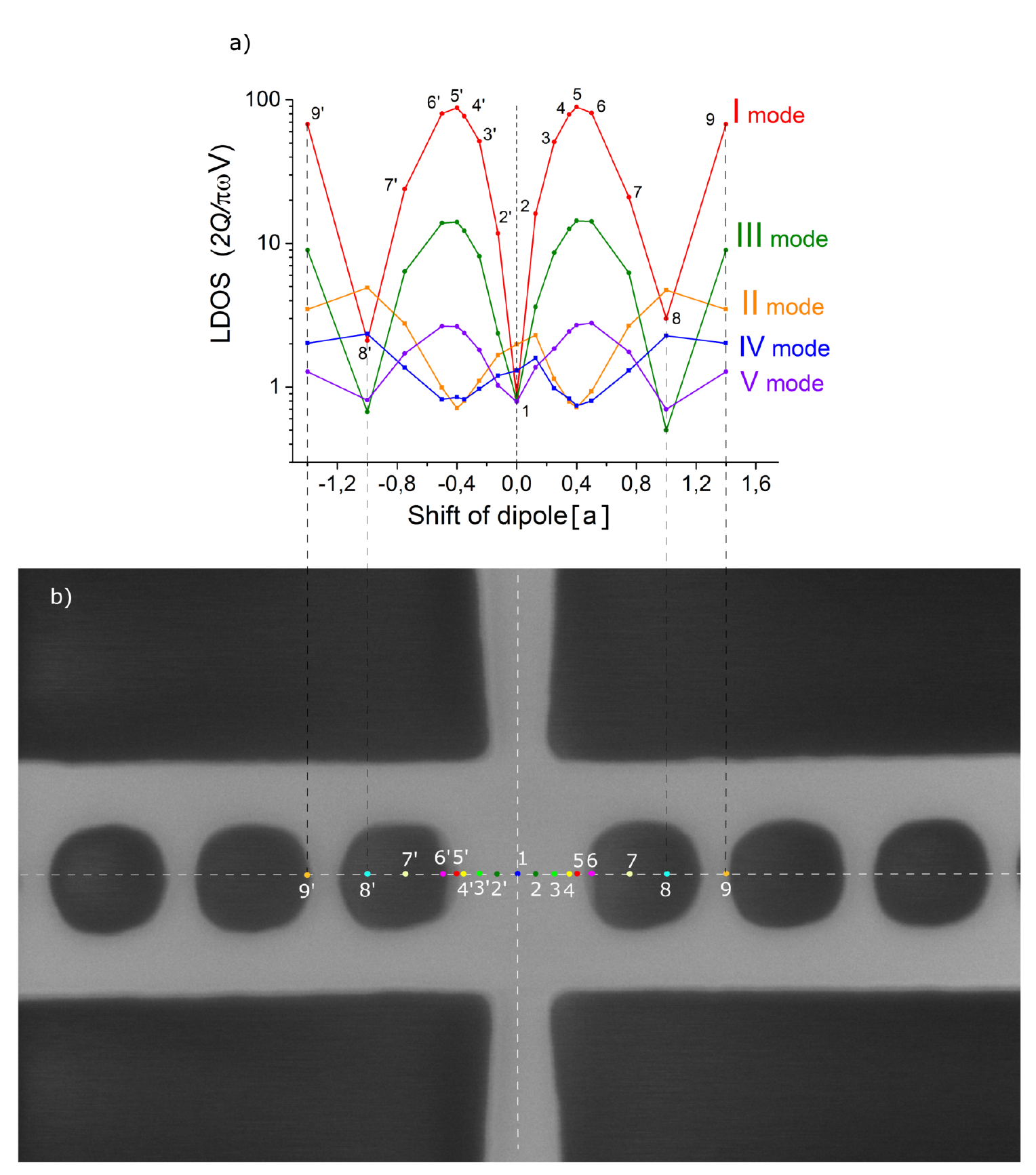}
\caption[]{(a) Simulated LDOS spectrum enhancement of the light emitted from a dipole placed on top of the freestanding cross-bar PhC cavity, where the dipole position is varied along the longitudinal direction ($x$-axis in Fig. \ref{fig:schematic_freestanding_PCC}). The positions are indicated in terms of the shift of the dipole from the center along the $x$-axis in Fig. \ref{fig:schematic_freestanding_PCC} and are weighted in units of $a$. The colorful curves correspond to the resonance modes labeled in roman numbers (I-V) in the plot. (b) SEM image of cavity region of the  fabricated and measured freestanding PhC cavity. The points indicate the position of the dipole for which the LDOS enhancement spectrum was computed in a) labeled with the corresponding numbers of the position. The PhC cavity has the following parameters: cavity length $195\nm$, quadratic tapered $N=18$ segments, periodicity $a=280\nm$.\\ 
Copyright declaration: First published in the PhD Thesis of Dr. Anna P. Ovvyan \lq{}\lq{}Nanophotonic circuits for single photon emitters" DOI: 10.5445/IR/1000093929}
\label{fig:simulated_ldos_spectrum_with_sem}
\end{figure}

According to the Purcell theory, strongly enhanced light-matter interaction with the emitter is provided when placing the source in the antinode of the electric field. Furthermore, the emitter's wavelength and polarization needs to match the resonance mode of the PhC cavity. Thus, the field pattern is the first indication where strong interaction with the emitter would occur. The Purcell enhancement factor approximates the single-resonant LDOS enhancement, where the latter one is proportional to the overlap integral between the electric field distribution of the resonance mode of the cavity and the electric pattern of the source. 3D-FDTD simulations are carried out using the MEEP software \cite{oskooiMeepFlexibleFreesoftware2010,tafloveAdvancesFDTDComputational2013} to compute the LDOS enhancement of an emitter placed on the cross-bar PhC cavity region. Therefore, the position ($x_0$,$y_0$,$z_0$) of a Gaussian pulse point-dipole -- with a major field component of $E_y$ -- on the cavity is varied and the corresponding excitation of the TE-like resonance modes is evaluated. Since the convolution of the electric field along the $x$-axis is the strongest, mainly the LDOS spectra dependence on the $x$-axis position of the dipole is considered. The relative LDOS enhancement spectrum is studied which is equal to the ratio of the LDOS with the cross-bar PhC cavity compared to the LDOS without any scattering elements (without PhC cavity). The normalized LDOS spectrum for different positions of the dipole on top of the cavity region along the $x$-axis is shown in Fig. \ref{fig:simulated_ldos_spectrum}. Again, these results are evaluated for a PhC cavity with a reduced number of mirror segments, than in the actual fabricated device, since the LDOS calculation needs to run for a full period of the cavity lifetime $2Q/f_{\mathrm{res}}$. The reduction of segments leads to a decrease of the computation time at a price of a decreased quality-factor, i.e. linewidth of the resonance peaks.\\
The spatial map of the simulated LDOS enhancement for odd and even resonance modes is illustrated in Fig. \ref{fig:simulated_ldos_spectrum_with_sem} and allows to sum up the results in a more conclusive way.
\begin{itemize}
\item The centered position of the emitter (position 1 in Fig. \ref{fig:simulated_ldos_spectrum_with_sem}) inhibits the coupling into odd modes, since there is a node in the electric field pattern of the corresponding odd resonance modes (Fig. \ref{fig:simulated_ldos_spectrum}) and at the same time provides only slight coupling to even resonance modes, due to a weak antinode in the corresponding electric field distribution. Thus, peaks corresponding to odd modes are missing in the LDOS enhancement spectrum of Fig.\ref{fig:simulated_ldos_spectrum} a).
\item Shifting the emitter away from the center of the cavity along the PhC leads to an increase in the enhancement for odd modes, while decreasing the coupling to even ones. The LDOS obtains a maximum for a shift of the emitter of $0.4a$ (position 5 in Fig. \ref{fig:simulated_ldos_spectrum_with_sem}) with respect to the symmetry plane of the center of the cavity, since the dipole is successfully coupled to the antinode of the corresponding odd resonance modes. Coupling into even modes is inhibited \cite{fehlerEfficientCouplingEnsemble2019}.
\item The LDOS enhancement factor of the coupling into odd modes is higher in comparison to even modes and is attributed to a higher $Q$-factor and also to a lower mode volume of odd modes for this particular PhC cavity.
\end{itemize}

\subsubsection*{Factors Impacting the LDOS Spectra.}
\begin{figure}
\includegraphics[width=\textwidth]{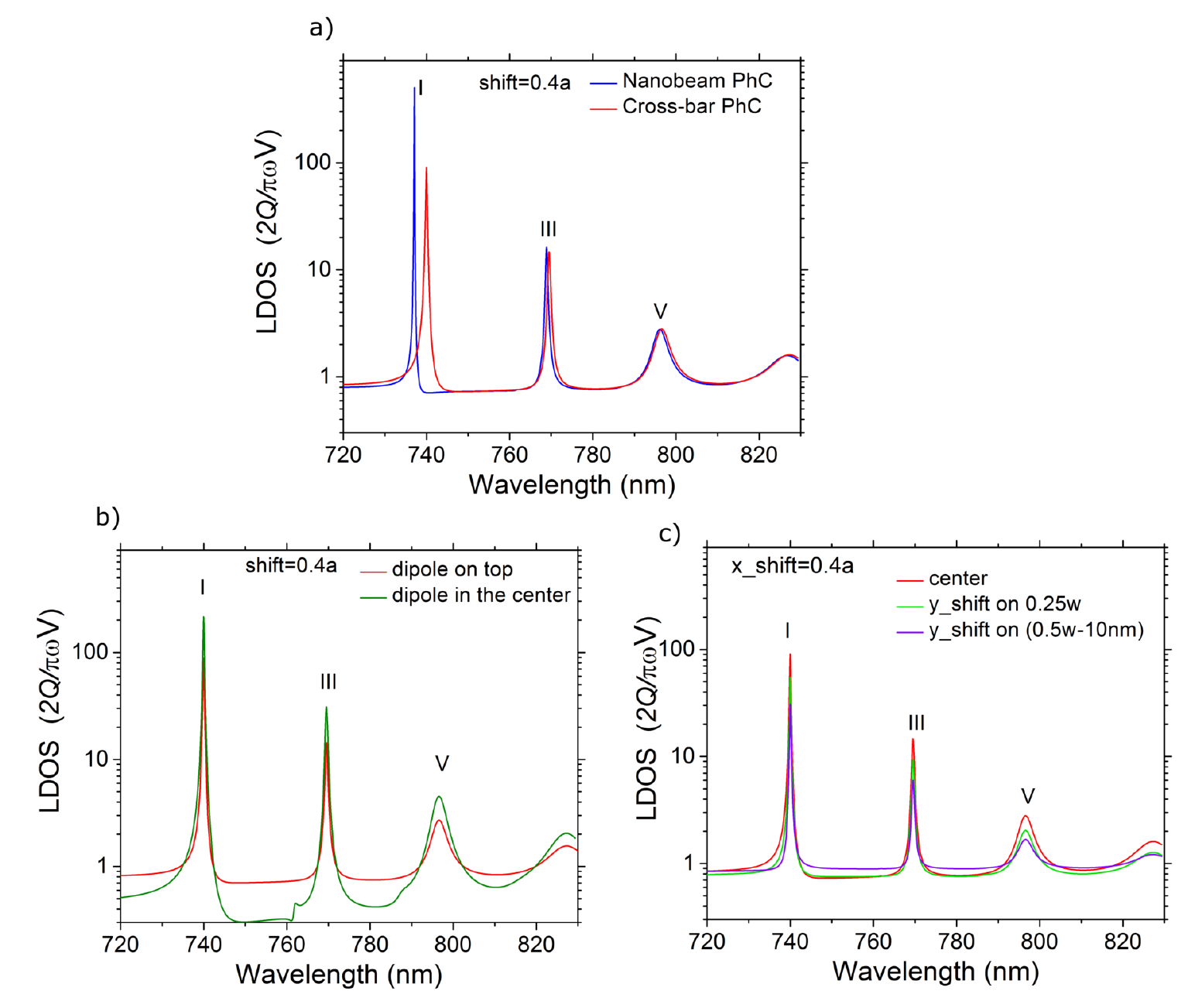}
\caption[]{(a) Simulated LDOS enhancement of emitted light from a dipole placed on top of the cavity region of a freestanding cross-bar (\textit{red curve}) and freestanding nanobeam (\textit{blue curve}) PhC cavity with identical parameters. (b) Simulated LDOS enhancement of emitted light from the dipole placed on top of cavity (\textit{red curve}) and inside the center of the cavity (\textit{green curve}) of a freestanding cross-bar PhC cavity. (c) Simulated LDOS enhancement of emitted light from the dipole placed on top of freestanding cross-bar PhC cavity in the center of the nanoguide (\textit{red curve}), shifted along $y$-direction on $0.25w=112.5\nm$ (\textit{light green curve}) and shifted along $y$-direction on $0.5w-10\nm=215\nm$ (\textit{violet curve}), where $w$ is the width of the waveguide. In all considered simulations of this figure the dipole was shifted to optimized distance $0.4a$ along the $x$-axis. Freestanding cross-bar PhC cavity has the following parameters: cavity length $195\nm$, quadratic tapered $N=18$ segments, periodicity $a=280\nm$.\\ 
	Copyright declaration: First published in the PhD Thesis of Dr. Anna P. Ovvyan \lq{}\lq{}Nanophotonic circuits for single photon emitters" DOI: 10.5445/IR/1000093929}
\label{fig:simulated_ldos_spectrum_on_top}
\end{figure}

The effect of the cross-bar on the LDOS is explored by 3D-FDTD simulations. The cross-bar mostly influences the fundamental resonance mode, which is expected from the mode's electric field distribution. In particular, it decreases the enhancement (about 5 times) and shifts the resonance wavelength (about $3\nm$) to longer wavelengths as shown in Fig. \ref{fig:simulated_ldos_spectrum_on_top} a). The influence of the cross-bar on the resonance conditions of the higher order modes is comparatively small.\\
The theoretically maximum enhancement factor of the spontaneous emission rate is obtained when the source is embedded inside the PhC cavity leading to an increase of the enhancement factor of all modes by $\approx 2$ times as shown in Fig. \ref{fig:simulated_ldos_spectrum_on_top} b). It is important to note, that another consequence of this optimized position is the suppression of the enhancement of the light coupled into the radiation modes due to the incomplete bandgap of the 1D PhC. The effect of the shift of the source from the optimized position along the $x$-axis is explained in the previous paragraph. While moving the source away from the center position along the width of the waveguide ($y$-axis) the LDOS decreases. A shift of $0.25w$ ($= 112.5\nm$, where $w$ - width of waveguide of $w=450\nm$) decreases the LDOS by 1.8 times. Placing the emitter at the edge of the waveguide (shift of the dipole of $0.5a-10\nm=215\nm$, violet curve) leads to a decrease of the LDOS enhancement of about 3 times in comparison with the central position.

\subsubsection*{Considerations Related to the Experimental Coupling of Single Photon Emitters to the Modes of the PhC Cavity.}
In conclusion, the maximal enhancement factor of the spontaneously emitted light coupled into odd resonance modes of the developed cross-bar PhC cavity is obtained in case of the $0.4a$ shifted dipole along the longitudinal direction from the symmetry plane in the center of the cavity (position 5 in Fig. \ref{fig:simulated_ldos_spectrum_with_sem}). The fabricated freestanding cross-bar PhC devices consist of $N=30$--$50$ mirror segments on each side of the cavity, leading to a drastic increase of the $Q$-factor and the LDOS enhancement factor in comparison with the computed values for a PhC cavity with $N=15$. It should be noted, that the LDOS enhancement spectra are computed for the case when the polarization of the source is perfectly aligned with the field of the TE-like resonance modes ($E_y$). However, in reality the discrepancy related with the positioning and the alignment of the emitter with respect to the polarization of the resonance modes will decrease observed enhancement factor.

\subsection*{The Coupling Efficiency of the Emitter into Resonance Modes of the Cross-Bar PhC Cavity Expressed by the $\beta$-Factor.}
\begin{figure}
\includegraphics[width=\textwidth]{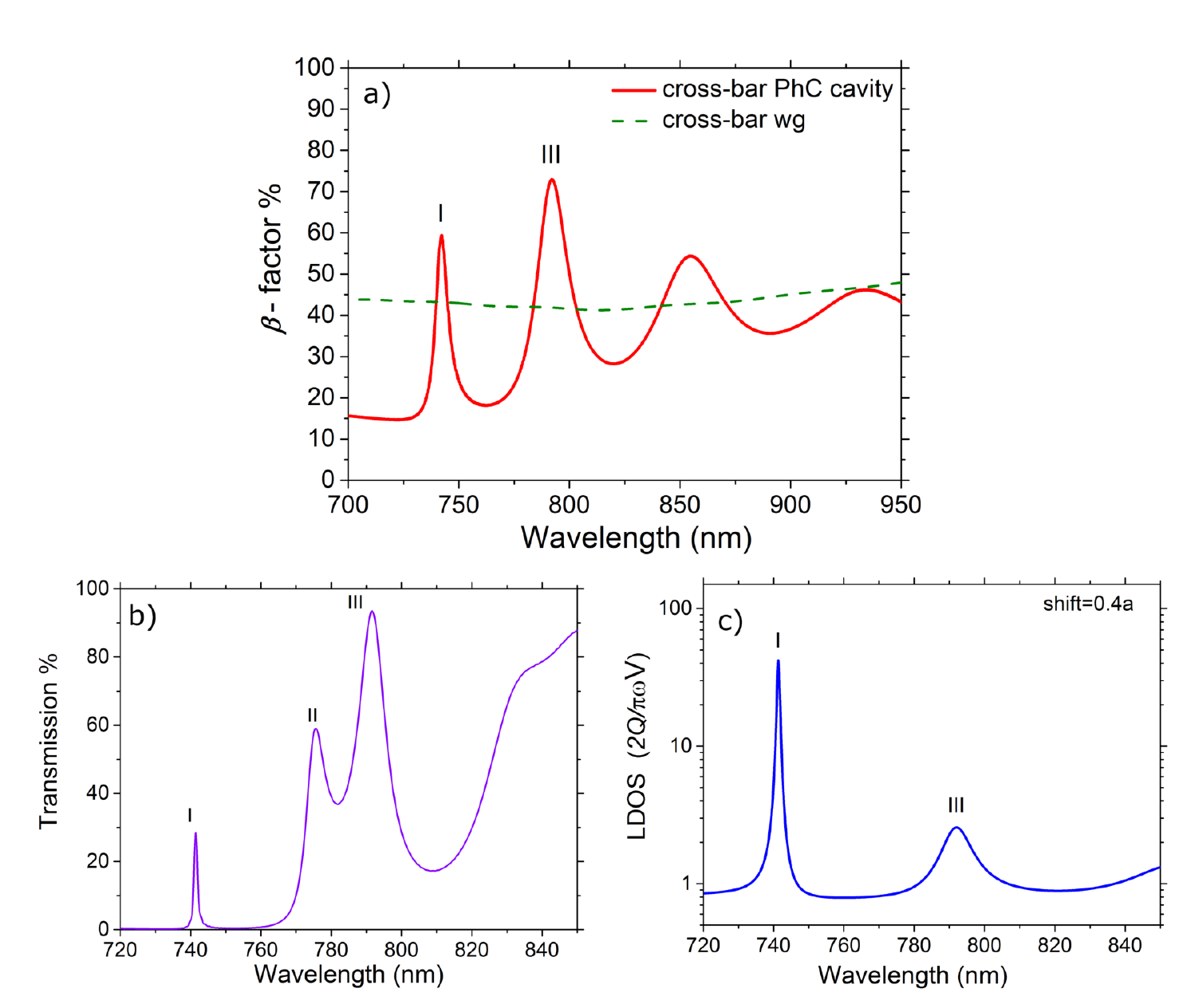}
\caption[]{(a) Simulated $\beta$-factor transmission coupling efficiency of emitted light into resonance modes and further longitudinal guided modes of freestanding cross-bar PhC cavity for the dipole positioned on top of the cavity region and shifted $0.4a$ away from the symmetry plane along the PhC. The dashed line shows the $\beta$-factor efficiency of coupling emitted light into waveguide mode for an identical geometry cross-bar freestanding waveguide without the PhC. (b) Simulated Transmission spectrum through the considered cross-bar PhC cavity. (c) Simulated LDOS enhancement spectrum of emitted light from the considered dipole. Simulated freestanding cross-bar PhC cavity has the following parameters: cavity length $195\nm$, $N=10$ segments, periodicity $a=280\nm$.\\ 
Copyright declaration: First published in the PhD Thesis of Dr. Anna P. Ovvyan \lq{}\lq{}Nanophotonic circuits for single photon emitters" DOI: 10.5445/IR/1000093929}
\label{fig:simulated_beta_factor}
\end{figure}

A main figure of merit for an emitter coupled to the cavity mode is the spectrally resolved $\beta$-factor. The coupling efficiency, expressed by the $\beta$-factor, correlates with the LDOS factor. Thus, it strongly depends on the spatial position of the emitter in the cavity region. The $\beta$-factor is the ratio between the amount of spontaneous emitted light (Poynting vector flux) coupled to the cavity and the total amount of spontaneous emitted light at this wavelength.
\begin{equation}
\beta=\frac{\Gamma_{\mathrm{PhC}}}{\Gamma_{\mathrm{PhC}}+\Gamma_{\mathrm{free}}}
\end{equation}
where $\Gamma_{\mathrm{PhC}}$ is the emission rate into the cavity mode and $\Gamma_{\mathrm{free}}$ is the free-space emission of the uncoupled emitter at the same wavelength. The $\beta$-factor is determined via 3D-FDTD simulations using the MEEP software, resulting in a wavelength-dependent $\beta$-factor. The wavelength-dependent coupling efficiency of the emitted light into the resonance modes of the freestanding cross-bar PhC and further into the waveguide is illustrated in Fig. \ref{fig:simulated_beta_factor} for a source placed in the optimal position (shift of $0.4a$ along the longitudinal direction) on the cavity . For comparison, the enhancement of the coupling efficiency on resonance is also computed into the waveguide mode of cross-bar waveguides with identical geometry but without the PhC (dashed green line in Fig. \ref{fig:simulated_beta_factor} a)).

\subsection{Hybrid Quantum Nanophotonics Established by Evanescent Optical Coupling}

\begin{figure}
\includegraphics[width=\textwidth]{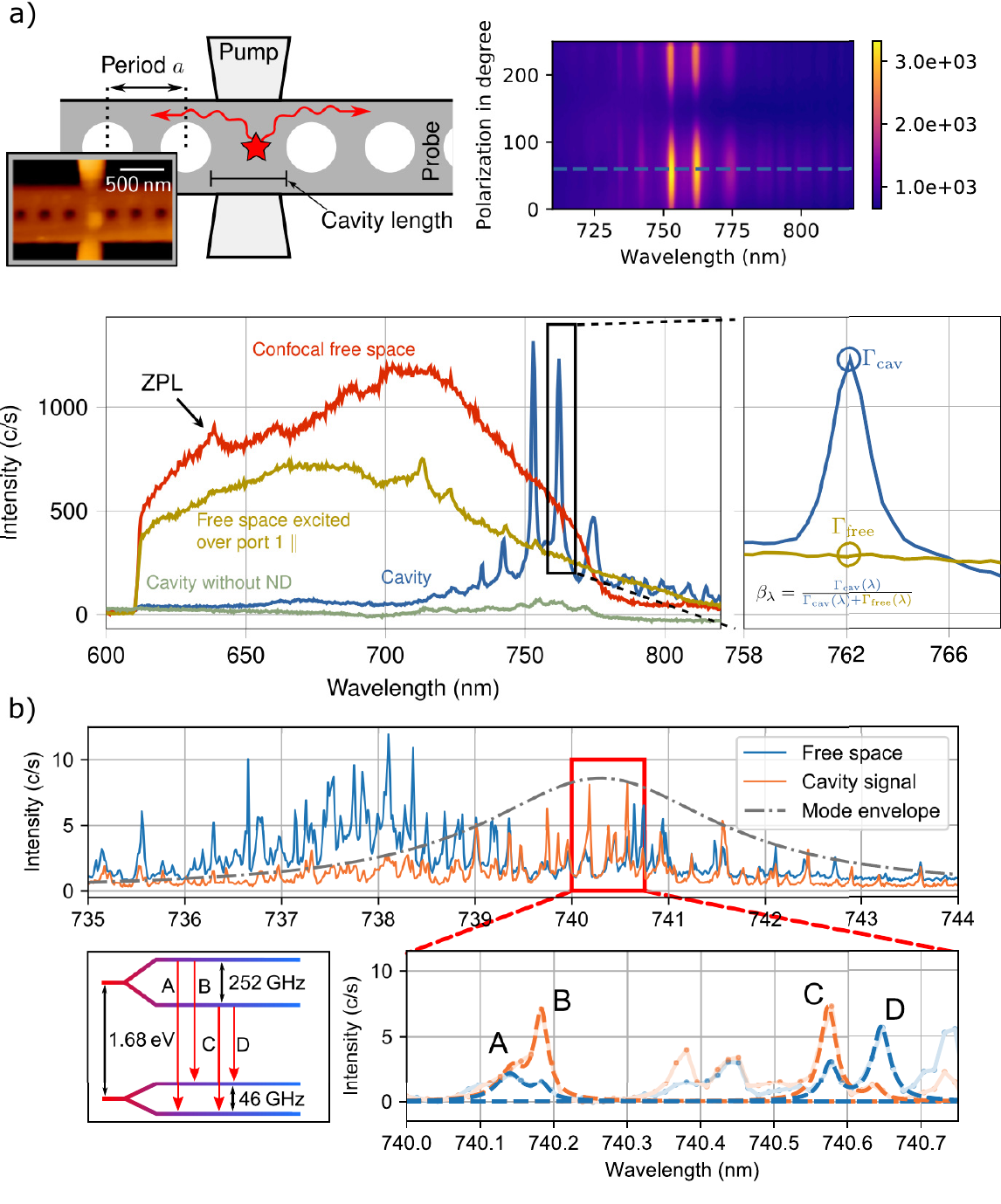}
\caption{\textbf{Hybrid quantum photonics based on evanescently coupled color centers in NDs.} (a) An ND containing an ensemble of NV center is pushed in the interaction zone of the cavity (\textit{upper left panel}). The fluorescence coupled to the cavity mode is measured and its polarization-dependence matches the polarization properties of the cavity-waveguide system (\textit{upper right panel}). The fluorescence signal is strongly enhanced when the system is resonant with the cavity modes yielding a $\beta _{\lambda}$-factor of 0.71 while the $\beta$-factor reaches 0.14 averaged over the full ensemble of NV centers. (b) Resonant coupling of the ZPL of SiV centers enables to resolve the fine-structure of individual centers. Individual optical transitions are strongly enhanced reaching Purcell-factors of 4.\\
Copyright declaration: (a) Reprinted with permission from Fehler et al., ACS Nano, 13(6) 6891-6898, 2019. Copyright (2019) American Chemical Society,
(b): Reprinted figure with permission from Fehler et al., Nanophotonics, 9(11), 3655-3662, 2020. Copyright (2020) author(s), licensed under
https://creativecommons.org/licenses/by/4.0/legalcode}
\label{fig:evanescent}
\end{figure}

Purcell enhancement of coherent photons can be achieved by resonantly enhancing the ZPL emission via evanescent coupling of a quantum emitter to photonic structures. In pioneering work, the ZPL of single NV centers was enhanced by placing the ND in the interaction region of microdisks \cite{santoriNanophotonicsQuantumOptics2010} and PhC cavities \cite{woltersEnhancementZeroPhonon2010,woltersCouplingSingleNitrogenvacancy2012}.
In this subsection, we summarize two recent developments on the evanescent coupling of color center in NDs to the mode of a high-Q, $\mathrm{Si}_3\mathrm{N}_4$  PhC cavity. First, we discuss the efficient coupling of an ensemble of NV center \cite{fehlerEfficientCouplingEnsemble2019}, as summarized in Fig. \ref{fig:evanescent} a). Second, we discuss the Purcell-enhanced emission of individual transitions of single SiV center \cite{fehlerPurcellenhancedEmissionIndividual2020}, as summarized in Fig. \ref{fig:evanescent} b). \\
On the one hand, $\mathrm{Si}_3\mathrm{N}_4$-photonics is a desired technology due to excellent photonic parameters such as low-photon loss. On the other hand, its strong background fluorescence prohibited its use in the quantum optics regime. In the first experiment \cite{fehlerEfficientCouplingEnsemble2019}, a cross-bar PhC cavity, as introduced in the previous section, was used to couple an ensemble of NV centers to individual cavity modes. In this configuration, the background fluorescence was successfully suppressed by about 20 dB. Based on that breakthrough the observation of an efficient coupling with a $\beta_\lambda$-factor of 0.71 was realized, meaning that more than 2/3 of the photons, which are spontaneously emitted at a wavelength of $762.1\nm$, are directed into the cavity mode. The overall $\beta$-factor reached 0.14 corresponding to one out of seven photons, which are spontaneously emitted at any wavelength, are coupled into the $\mathrm{Si}_3\mathrm{N}_4$-waveguide. This corresponds to an overall Purcell-factor  $F_P=\beta / (1-\beta)$ of 0.16 which is a reasonably promising number given that it corresponds to the enhancement averaged over all emitters in the ensemble. \\
The fact that individual emitters, which have a dipole well aligned with the cavity mode, can have a much higher Purcell-enhancement was demonstrated in successive work \cite{fehlerPurcellenhancedEmissionIndividual2020}. NDs, now containing SiV center, were pushed into the interaction zone of the PhC cavity. Overall, a Purcell enhancement of more than 4 on individual optical transitions was achieved, in line with four out of five spontaneously emitted photons guided into the photonic device. Also, the finestructure of individual SiV center was observed within the mode envelope. To reach this goal the overlap between the SiV center's dipole and the cavity electric field was repeatedly optimized by AFM nanomanipulation. Furthermore, the PhC resonance condition was tuned to achieve spectral overlap of the emitter ZPL and the cavity mode. However, the expected lifetime-shortening due to the Purcell effect to approximately 340 ps was not yet observed.  

\subsection{Hybrid Quantum Nanophotonics Established by Embedded Optical Coupling}

\begin{figure}
\includegraphics[width=\textwidth]{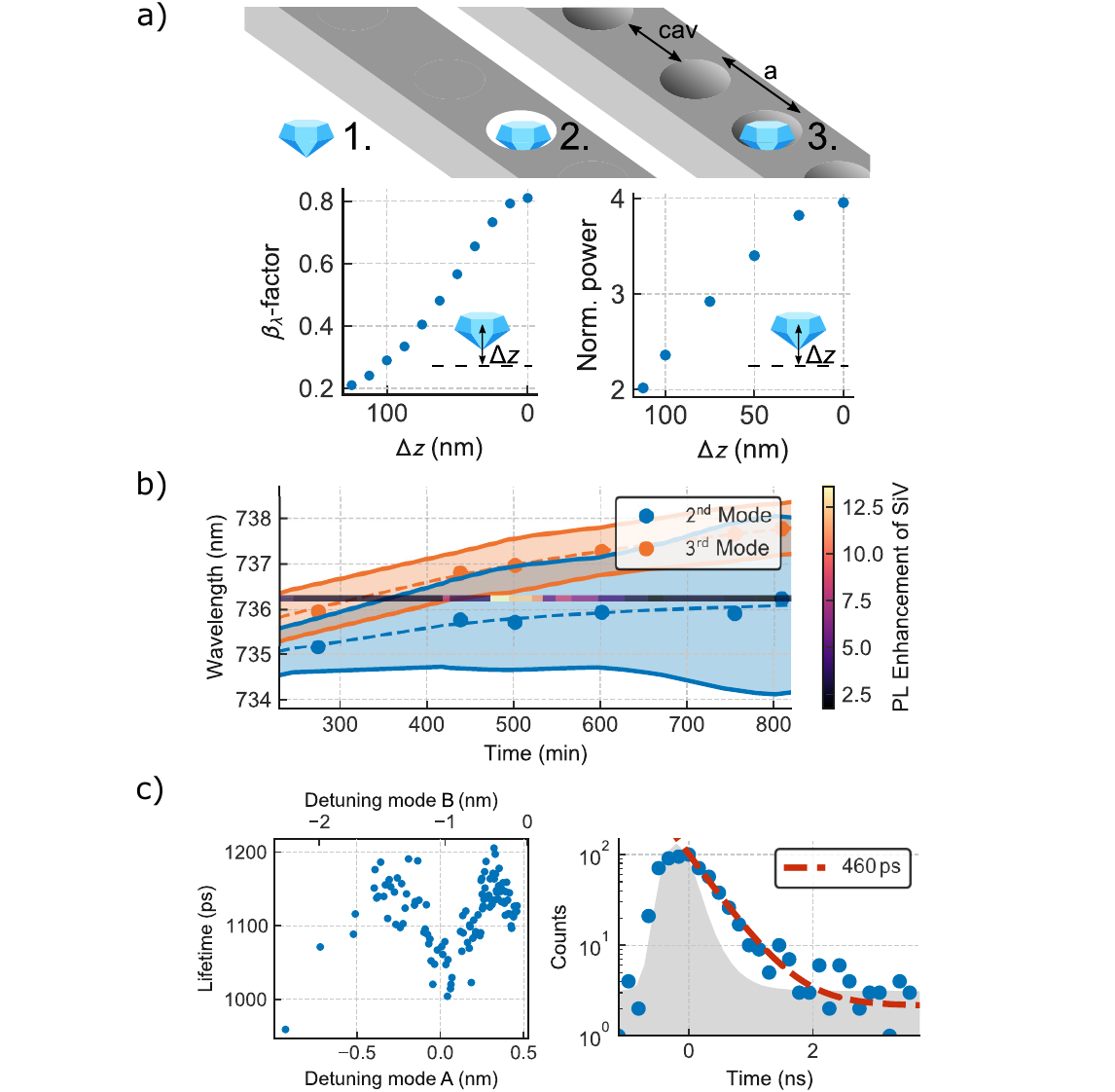}
\caption[]{\textbf{Hybrid quantum photonics based on embedded optical coupling.} (a) An ND containing SiV center is positioned with high precision in the hole of a PhC cavity. Three cases are compared to demonstrate the advantage of embedded coupling. 1. The ND in free space. 2. The ND placed inside one hole of a bare waveguide, studying only the effect of waveguiding. 3. The ND placed in the hole of a PhC cavity where signal enhancement arises from waveguiding together with Purcell-enhancement (\textit{upper panel}). The simulated improvement of the $\beta_{\lambda}$-factor and the normalized power is plotted as function of the displacement of the ND from the center of the waveguide (\textit{lower panel}). (b) The experimentally determined PL-enhancement is shown as function of the detuning from two modes. The strongest enhancement of more than 14-times is reached when the optical transition is in resonance with the composite mode, comprised of coupling to both, the $2^{nd}$ and $3^{rd}$, mode. (c) Lifetime shortening for the next order composite mode is plotted as function of detuning (\textit{left panel}). Significant lifetime reduction is reached when the optical transition is in resonance with the cavity mode. The shortest recorded lifetime reached $460\,\mathrm{ps}$, close to the resolution limit of the detection system marked as gray shaded area (\textit{right panel}).\\
Copyright declaration: (a--c) Reprinted with permission from Fehler et al., ACS Photonics 8(9), 2635–2641, 2021. Copyright (2021) author(s), licensed under https://creativecommons.org/licenses/by-nc-nd/4.0/legalcode}
\label{fig:embedded}
\end{figure}

The light-matter coupling strength can be further increased by optimizing the overlap of the cavity field with the position of the quantum emitter. Ideally, the emitter is embedded directly into the photonics material at the position of the mode field maximum. With respect to color centers in diamond, this means that the nanophotonics material needs to be diamond, capable to host the color centers. Engineering photonic devices out of diamond for that purpose was pioneered more than a decade ago. The ZPL of NV centers was resonantly enhanced by fabrication ring resonators in diamond \cite{faraonResonantEnhancementZerophonon2011} and PhC cavities in monocrystalline diamond \cite{faraonCouplingNitrogenVacancyCenters2012}. Integrated nanophotonics was realized by coupling single NV center to all-diamond ring resonators which are coupled to all-diamond waveguides that include grating couplers \cite{hausmannIntegratedDiamondNetworks2012}. The research field rapidly evolved over the following years and is nowadays among the most promising platforms for the realization of quantum applications such as quantum repeater and quantum networks \cite{bhaskarExperimentalDemonstrationMemoryenhanced2020}. Recently, suspended PhC cavities in polycrystalline diamond were engineered to reach 2.5-fold ZPL enhancement of SiV center. Here, the spectral overlap between the SiV ZPL and the cavity mode was realized without the need of any post-processing taking advantage of the non-homogeneous thickness of the diamond layer \cite{ondicPhotonicCrystalCavityenhanced2020}. The high precision of nanomanipulation, as discussed in the previous sections, now makes embedded coupling also accessible for hybrid quantum systems.\\
The idea of hybrid embedded coupling is, that instead of placing the ND outside the photonic device it is integrated into the photonics material. The photonics material itself does not need to be diamond, in contrast to the all-diamond devices. Hybrid embedded coupling can, for example, be achieved by establishing pre-defined nanopockets \cite{alagappanDiamondNanopocketNew2018,frochPhotonicNanobeamCavities2020}. Therefore, a very precise nanomanipulation is required. In the work \cite{fehlerHybridQuantumPhotonics2021} the precision of the QPP was pushed to an extreme by placing a ND, which contains SiV centers, inside the hole of a PhC cavity by using the AFM-based pick and place technique. As a result, the corresponding coupling strength got another boost from an improved Purcell-enhancement and, at the same time, from waveguiding effects. By employing further optimization steps, such as two-mode coupling and precise cavity-resonance tuning, the resulting photon flux was increased by 14-times as compared to free space. For the first time, also the corresponding lifetime shortening to below 460 ps was directly observed, only limited by the technical time-resolution of the setup. The results are summarized in Fig. \ref{fig:embedded}. Therefore, the hybrid quantum photonics platform based on embedded coupling is among the few spin-based photonics platforms with an operation bandwidth beyond GHz-rates. The platform is now at a stage to strive the realization of spin-based quantum applications.


\section{Summary}

In this chapter, we summarized the performance of hybrid quantum photonics based on $\mathrm{Si}_3\mathrm{N}_4$-devices postprocessed with NDs containing color centers, in particular NV and SiV-center. We discuss scalable, low-loss, integrated, $\mathrm{Si}_3\mathrm{N}_4$-photonics elements which are necessary for the realization of modern, hybrid quantum photonics on-chip \cite{lombardiPhotostableMoleculesChip2018,elshaariHybridIntegratedQuantum2020,khasminskayaFullyIntegratedQuantum2016,wangIntegratedPhotonicQuantum2020,pyatkovCavityenhancedLightEmission2016,khasminskayaWaveguideIntegratedLightEmittingCarbon2014,thomsonHighPerformanceMach2013,pelucchiPotentialGlobalOutlook2022,wanLargescaleIntegrationArtificial2020,uppuQuantumdotbasedDeterministicPhoton2021,machielseQuantumInterferenceElectromechanically2019} and programmable photonic circuits \cite{feldmannAllopticalSpikingNeurosynaptic2019,feldmannParallelConvolutionalProcessing2021,bogaertsProgrammablePhotonicCircuits2020,dongHighspeedProgrammablePhotonic2022,borregaardOneWayQuantumRepeater2020}. Designed, fabricated, experimentally and computationally fully investigated 1D, freestanding, cross-bar PhC cavities integrated on chip allow us to simultaneously solve three outstanding challenges, namely: 
\begin{enumerate}
\item 
Individual, optical transitions of the emitter can efficiently be coupled to the high-quality, small-volume PhC cavity resonance modes.
\item 
The established pump-probe design enables simultaneous collection of enhanced fluorescence through the cavity-waveguide channel, while the excitation of the emitter is realized via crossed waveguides. Thus, both pump and read-out ports are on-chip and are accessible for additional on-chip photonic technology.
\item 
The cross-bar design ensures spatial separation of the excitation and the enhanced emission of the coupled emitter in the cavity region, enabling strong suppression ($\approx -20\,\mathrm{dB}$) of the pump waveguide cross-talk and correspondingly strong suppression of background fluorescence. 
\end{enumerate}
Experimentally efficient integration of NV and SiV centers into the developed cross-bar PhC cavities and enhancement of the spontaneous emission rate is demonstrated in the works \cite{fehlerEfficientCouplingEnsemble2019,fehlerPurcellenhancedEmissionIndividual2020,fehlerHybridQuantumPhotonics2021}. We explore the LDOS enhancement spatial map and determine an optimized position for the emitter in the cavity region in order to obtain a maximum enhancement of the emission coupled into certain resonance modes. The insights were taken into account during the positioning of the NDs into the cavity interaction region. The simulated coupling efficiency, evaluated coupling strengths and enhanced spontaneous emission rate of the photon source in resonance with the PhC and with further waveguide modes of the integrated photonic device resulted in simulated $\beta$-factors of  $\beta_\lambda=75\%$, close to the experimental results of $\beta_\lambda = 81\%$ \cite{fehlerPurcellenhancedEmissionIndividual2020}. \\
We elaborate on the potential of the platform in quantum photonics and as spin-photon interface. Our hybrid device enables coherent optical control with GHz-bandwidth, access to long-lived electron- and nuclear spin as well as on-chip control. Ongoing device optimization is required to unfold the full potential of the platform and includes the following steps: 
\begin{itemize}
\item 
\textbf{Improve the Q-factor of the hybrid photonic devices.} In particular, the high-Q of the bare PhC needs to persist the QPP. For example, when placing a ND in the interaction zone towards evanescent coupling the Q-factor dropped from $Q_{\mathrm{bare}}=1000$ for the bare cavity to $Q_{\mathrm{hybrid}}=480$ for the established hybrid platform \cite{fehlerPurcellenhancedEmissionIndividual2020}. In order to reach higher Q-factors, the scattering induced by the NDs needs to be reduced. Consequently, much higher Purcell enhancement beyond 500 becomes feasible.
\item
\textbf{Establish color centers in tailored NDs.} One path of reducing the scattering induced by the NDs is to create color centers with good spectral and spin properties in tailored NDs. Tailoring could mean to use much smaller NDs on the order of 10 nm or NDs with controlled shape. This challenging task requires further developments on the material side. Controlling the size and shape of the ND not only reduces the amount of scattering. It also makes the influence arising from introducing a particle with different refractive index into the photonic device more predictable. Therefore, the photonic designs become more reliable and can be optimized for best performance of the hybrid device instead of the bare photonic device. For example, system imperfections such as scattering or a changed refractive index could be taken into account.
\item 
\textbf{Optimize all degrees of freedom of the coupling term.} Optimizing all degrees of freedom of the coupling term (spatial, spectral and polarization alignment) unfolds the full potential of the quantum photonic devices. In particular, the positioning and dipole alignment can, in principle, be fully controlled by means of nanomanipulation. Therefore, the overall system cooperativity can be pushed to its theoretical limits.
\item 
\textbf{Speed-up the QPP towards a high-throughput technology.} In order to make the complete fabrication chain efficient, current bottlenecks in terms of fabrication time need to be speed up. In particular, the QPP is a slow process and needs a boost in processing time in order to be compatible with the large-scale fabrication of $\mathrm{Si}_3\mathrm{N}_4$-photonics. Ideas to improve large-scale integration include lithographically positioned color centers into photonic integrated circuits \cite{schrinnerIntegrationDiamondBasedQuantum2020} or emitter precharacterization supported by machine learning \cite{kudyshevRapidClassificationQuantum2020} as discussed in this chapter.
\end{itemize}
We believe that hybrid quantum photonics based on $\mathrm{Si}_3\mathrm{N}_4$-photonics and color centers in NDs is a promising route to establish quantum photonics in a low-loss, high-throughput, on-chip platform. 
Essential elements of such integrated photonic circuits include tunable filters. Cascaded MZI tunable filters equipped with microheaters on both arms of the MZI to provide a thermo-optical effect are discussed in this chapter. The filter enables to transmit a desired wavelength at the interference maximum and, at the same time, blocks the excitation light at the interference minimum. The developed spiral-type micro-heaters ensure the lowest switching power of $P_\pi=12.2\mW$ in comparison with meander-shaped heaters, which compare favorably with previous reported results for $\mathrm{Si}_3\mathrm{N}_4$-platforms. Designed tunable filters are multifunctional and enable to transmit and block the desired wavelengths in a wide wavelength range. Therefore, such circuits can be used in a large variety of future applications in hybrid quantum photonics. We show the transmission of light at $\lambda=740\nm$, according to the SiV fluorescence, and the corresponding suppression of the excitation light at $\lambda=532\nm$, where cascading two stages of the MZI improves the filtration depth to obtain $\mathrm{ER}=34.7$--$36.5\,\mathrm{dB}$. The complete control on-chip will enable to scale the platform to many qubits interconnected by many photonic channels which are operated with quantum states of light. At this point the full capabilities of $\mathrm{Si}_3\mathrm{N}_4$-photonics can be exploited. \\
The established coupling and strong Purcell-enhancement of individual transitions of color centers in NDs is a prerequisite for on-chip quantum optics experiments. The efficient funneling of single photons into the photonic waveguides enables in the near future to route those photons through complex photonic structures. Operating the platform in the quantum regime enables the realization of emerging quantum optics applications in an integrated fashion \cite{leeIntegratedSinglePhoton2020}. Applications include long-distance quantum communication and on-chip quantum computing \cite{kokLinearOpticalQuantum2007} as well as quantum photonic networks \cite{ozaydinDeterministicPreparationStates2021}. Furthermore, inherent nuclear spins within the diamond host residing in close proximity to the color center could be deployed as long-lived quantum memory. Such storage of quantum information would prove useful as a photonic memory \cite{kokLinearOpticalQuantum2007} or in applications like quantum repeaters \cite{munroQuantumRepeaters2015} or quantum networks \cite{bhaskarExperimentalDemonstrationMemoryenhanced2020,rufQuantumNetworksBased2021} as well as spin-based quantum technology in general \cite{awschalomQuantumTechnologiesOptically2018,atatureMaterialPlatformsSpinbased2018}.

\appendix

\section*{Appendix}

\section*{Acknowledgements}

The summarized projects were funded by the Baden-Württemberg Stiftung in project Internationale Spitzenforschung, the Deutsche Forschungsgemeinschaft (DFG, German Research Foundation) in Project No. 398628099, the BMBF and VDI in Verbundsproject Q.Link.X and QR.X. AK acknowledges support from IQst.

\bibliographystyle{spphys}
\bibliography{HybridQuantumChapter}

%

\end{document}